\documentclass[prd,a4paper,twocolumn,preprintnumbers,groupedaddress,nofootinbib]{revtex4}
\pdfoutput=1
\usepackage{amsmath}
\usepackage{graphicx}
\usepackage{xspace}
\usepackage{color}
\usepackage{units}
\usepackage{slashed}
\usepackage[hyperfootnotes=false]{hyperref}

\def \L {\mathcal{L}} 

\def \epsilon {\varepsilon} 

\newcommand{\Br}{\text{Br}}
\newcommand{\del}{\partial}
\newcommand{\dd}{\mathrm{d}}
\newcommand{\oh}{\nicefrac{1}{2}}
\newcommand{\tth}{\nicefrac{3}{2}}

\begin{document}

\title{Indirect searches of dark matter via polynomial spectral features}

\preprint{ULB-TH/16-07}

\author{Camilo Garcia-Cely}
\email{Camilo.Alfredo.Garcia.Cely@ulb.ac.be}
\affiliation{Service de Physique Th\'eorique, Universit\'e Libre de Bruxelles, Boulevard du Triomphe, CP225, 1050 Brussels, Belgium}

\author{Julian \surname{Heeck}}
\email{Julian.Heeck@ulb.ac.be}
\affiliation{Service de Physique Th\'eorique, Universit\'e Libre de Bruxelles, Boulevard du Triomphe, CP225, 1050 Brussels, Belgium}

\hypersetup{
    pdftitle={
		Indirect searches of dark matter via polynomial spectral features},
    pdfauthor={Camilo Garcia-Cely, Julian Heeck}
}


\begin{abstract}
We derive the spectra arising from non-relativistic dark matter annihilations or decays into intermediary particles with arbitrary spin, which subsequently produce neutrinos or photons via two-body decays. Our approach  is model independent and predicts spectral features restricted to a kinematic box. The overall shape within that box is a polynomial determined by the polarization of the decaying particle. We illustrate our findings with two examples. First, with the neutrino spectra arising from dark matter annihilations into the massive Standard Model gauge bosons. Second, with the gamma-ray and neutrino spectra generated by dark matter annihilations into hypothetical massive spin-2 particles. 
Our results are in particular applicable to the 750 GeV diphoton excess observed at the LHC if interpreted as a spin-0 or spin-2 particle coupled to dark matter. We also derive limits on the dark matter annihilation cross section into this resonance from the non-observation of the associated gamma-ray spectral features by the H.E.S.S.~telescope.
\end{abstract}

\maketitle


\section{Introduction}

There is plenty of evidence for the existence of particles beyond the Standard Model (SM)~\cite{Bertone:2010zza,Bergstrom:2000pn, Bertone:2004pz}. These particles, which constitute the so-called dark matter (DM),  have only been observed through their gravitational interactions and, consequently, very little is known about  their mass or quantum numbers. Currently, there are at least three search strategies aiming to identify those properties~\cite{Bergstrom:2012fi,Profumo:2013yn}. One of them, indirect searches of DM, relies on the assumption that DM could annihilate or decay into SM particles, a common feature of the popular weakly interacting massive particle (WIMP) DM. 
Of particular importance are photon and neutrino final states because these particles are not deflected by electromagnetic fields on their way to Earth and hence point directly back to the place where they were produced. This allows us to focus our search on regions of the sky where the concentration of DM is known to be high.

Still, one of the biggest challenges to this endeavor is the proper identification of the astrophysical backgrounds. Because of that, only DM annihilations or decays leading to sharp spectral features that can stand out over the featureless soft background are considered smoking-gun signatures for discovery.  Such spectral features can be classified into three categories: monochromatic lines, virtual internal Bremsstrahlung (VIB) and box-shaped spectra (see Ref.~\cite{Bringmann:2012ez} for a review). The former arise in two-body decays or annihilations into neutrinos or photons.   
The gamma-ray spectral feature associated to VIB  takes place, for instance, when a symmetry of the DM initial state is not fulfilled by a particular two-body final state but it is satisfied by the same final state once a  photon is emitted from the internal lines~\cite{Bringmann:2007nk}. 
Similar features can occur for neutrinos if they belong to a three-body final state~\cite{Aisati:2015ova}.
Finally, box-shaped spectra appear when DM annihilates or decays into scalar mediators, which subsequently decay into neutrinos or photons~\cite{Ibarra:2012dw}. All of these spectra can mimic lines and can in particular be easily disentangled from a featureless soft background. 

In this article, we will consider box-shaped spectra for mediators with arbitrary spin. The resulting spectra are derived model-independently and are determined only by the polarization of the intermediate particle decaying into neutrinos or photons. In particular, when the intermediate particle is unpolarized, we will show that a flat box-shaped spectrum is generated. 

We will use these concepts to describe the neutrino spectrum generated from DM annihilation into polarized SM gauge bosons. Then we will apply them to DM annihilations into massive spin-2 particles. 
This discussion is timely given the recent \unit[750]{GeV} diphoton excess at the LHC~\cite{CMS:2015dxe,CMS:2016owr,atlas_diphoton,ATLAS_diphoton_moriond}, which can only be interpreted as a new spin-0 or spin-2 resonance according to the Landau--Yang theorem~\cite{Landau,Yang:1950rg}. While not yet statistically significant, hundreds of articles have already been written about possible interpretations and implications of this excess.
If confirmed, the most pressing issue would be the determination of the new particle's spin via the angular diphoton distribution, followed by measurements of couplings and branching ratios. For spin-2 resonances, a common gravity-inspired model benchmark is to assume universal couplings to all SM particles via the total energy--momentum tensor, implying fixed branching ratios. This scenario is however disfavored by dilepton searches~\cite{Franceschini:2015kwy,Low:2015qep,Giddings:2016sfr}, making different couplings to photons and leptons necessary to fit the diphoton resonance, which can be obtained for example in generalized Randall--Sundrum~\cite{Randall:1999ee} models~\cite{Davoudiasl:2000wi,Falkowski:2016glr,Hewett:2016omf}.

It is not far fetched to assume that the diphoton resonance $R$ also couples to DM~\cite{Chu:2012qy,Backovic:2015fnp,Ellis:2015oso,Han:2015cty}. Together with the name-giving decay $R\to\gamma\gamma$, this implies interesting indirect-detection signatures, e.g.~monochromatic lines from $\text{DM DM}\to R^*\to \gamma\gamma$~\cite{Chu:2012qy,Park:2015ysf,D'Eramo:2016mgv,Ge:2016xcq} or box-shaped spectra from $\text{DM DM}\to R R\to 4\gamma$~\cite{Chu:2012qy,Choi:2016cic}. As we will show in this article, the gamma-ray spectrum of the latter depends on the polarization of the diphoton resonance $R$. We also consider the potential spectral features of neutrinos, even though the diphoton resonance has no dilepton counterpart as of now.

This article is organized as follows. We start Sec.~\ref{sec:Wbosondecay} with a detailed discussion of DM annihilation into intermediary (polarized) $W$ bosons, followed by the decay into neutrinos, to illustrate the resulting spectral features. In Sec.~\ref{sec:decay}, we generalize this example to intermediary particles of arbitrary spin and present the possible polynomial box-shape features from their decay. Sec.~\ref{sec:production} gives a number of examples on how to produce a polarized spin-2 particle from DM annihilation and the corresponding gamma-ray and neutrino spectra.
Motivated by these examples, we discuss model-independent spectra from a spin-2 mediator coupled to DM in Sec.~\ref{sec:spectra} and discuss the connection to the diphoton resonance in Sec.~\ref{sec:diphoton}. Finally, we conclude in Sec.~\ref{sec:conclusion}. 

\section{Annihilating DM into polarized gauge bosons}
\label{sec:Wbosondecay}

We start off with a simple  example to illustrate the concepts and familiarize ourselves with the notation. Consider WIMP DM that annihilates (or decays) into $W$ or $Z$ bosons.
If the DM particle is much heavier than the electroweak scale, the massive gauge bosons arising from DM annihilations are expected to be polarized~\cite{Drees:1992am}. This can be understood with the Goldstone-boson equivalence theorem (GBET)~\cite{Cornwall:1974km,Lee:1977eg}: the longitudinal modes are associated to the Goldstone boson that is absorbed by the vector particle in order to gain mass and are thus related to the interactions of the SM scalar doublet $H$, whereas the transverse modes arise by pure gauge interactions. Since scalar and gauge interactions may have different strengths, the emission of longitudinal and transverse bosons occurs at different rates.\footnote{ A similar situation takes place in the top-quark decay $t\to b W$, in which the polarization of the $W$ boson is dominantly \emph{longitudinal}.}

In order to illustrate this more precisely, let us consider models where WIMPs with non-relativistic velocities annihilate into $W^+W^-$. We find it convenient to introduce the branching ratio into the different polarizations $m$ and $n$ of the $W$ bosons as
\begin{align}
\text{Br}_{mn} \equiv \frac{\sigma v \left( \text{DM DM} \to W^+_m W^-_n\right)}{{\displaystyle\sum_{m'n'}}\sigma v \left( \text{DM DM} \to W^+_{m'} W^-_{n'}\right)} \,.
\end{align}
Here, $m=0$ corresponds to longitudinal and $m=\pm 1$ to transverse modes relative to the $W_m$ momentum. In models where DM is a Majorana fermion with $SU(2)_L$ quantum numbers (examples could be Wino, Higgsino~\cite{Thomas:1998wy} or Minimal DM~\cite{Cirelli:2005uq,Cirelli:2009uv,Cirelli:2007xd}), we find
\begin{align}
\text{Br} = \begin{pmatrix} \frac{1}{2}& 0 & 0 \\ 0 & 0 & 0 \\ 0 & 0 & \frac{1}{2} \end{pmatrix} +  {\cal O} \left(\tfrac{M_W^2}{M_\text{DM}^2}\right) .
\label{eq:Pwino}
\end{align}
This means that the gauge bosons are mostly transverse when the DM is much heavier than the $W$~\cite{Drees:1992am}. This is not necessarily the case for scalar DM: the simplest example corresponds to the situation when DM is an $SU(2)_L$ singlet $S$~\cite{Silveira:1985rk,McDonald:1993ex}. Then, the only non-trivial interaction of  DM with the SM takes place via the so-called Higgs portal $S^2 |H|^2$. In that case, we find
\begin{align}
\text{Br} = \begin{pmatrix} 0& 0 & 0 \\ 0 & 1 & 0 \\ 0 & 0 & 0 \end{pmatrix} + {\cal O} \left(\tfrac{M_W^2}{M_\text{DM}^2}\right) ,
\label{eq:Psinglet}
\end{align}
which implies that the gauge bosons are mostly longitudinally polarized, easily understood with help of the GBET. 
An intermediate situation is the case of the Inert Doublet model~\cite{Deshpande:1977rw,Barbieri:2006dq,Ma:2006km,LopezHonorez:2006gr}, in which DM is described by a scalar $SU(2)_L$ doublet $H_\text{DM}$ that also interacts with the SM scalar via quartic couplings. Of particular interest for the annihilation into $W^+W^-$  is the Higgs portal coupling $\lambda_3 |H_\text{DM}|^2 |H|^2$. The branching ratios now depend on both gauge and quartic couplings~\cite{Garcia-Cely:2015khw},
\begin{align}
\text{Br} = \tfrac{1}{2( g^4+2\lambda_3^2)} \begin{pmatrix} g^4& 0 & 0 \\ 0 & 4\lambda_3^2 &  0 \\ 0 & 0 & g^4 \end{pmatrix} +  {\cal O} \left(\tfrac{M_W^2}{M_\text{DM}^2}\right) .
\label{eq:PIDM}
\end{align}
For large values of $\lambda_3$, scalar interactions dominate the DM dynamics, the gauge bosons are longitudinally polarized and we recover Eq.~\eqref{eq:Psinglet}. In contrast, for negligible quartic couplings, the gauge interactions determine the properties of DM, the gauge bosons produced in annihilations are mostly transverse and we recover Eq.~\eqref{eq:Pwino}.
Obtaining unpolarized $W$ bosons requires tuning the quartic coupling so that $4\lambda_3^2\simeq  g^4$.

We thus generically expect gauge bosons produced in TeV DM annihilations to be polarized as a result of the GBET. The purpose of this work is to investigate the consequences of such polarization  on the spectrum arising from the decay of the gauge bosons, and to generalize this result to intermediate particles of arbitrary spin.  

\begin{figure}[t]
\includegraphics[width=0.45\textwidth]{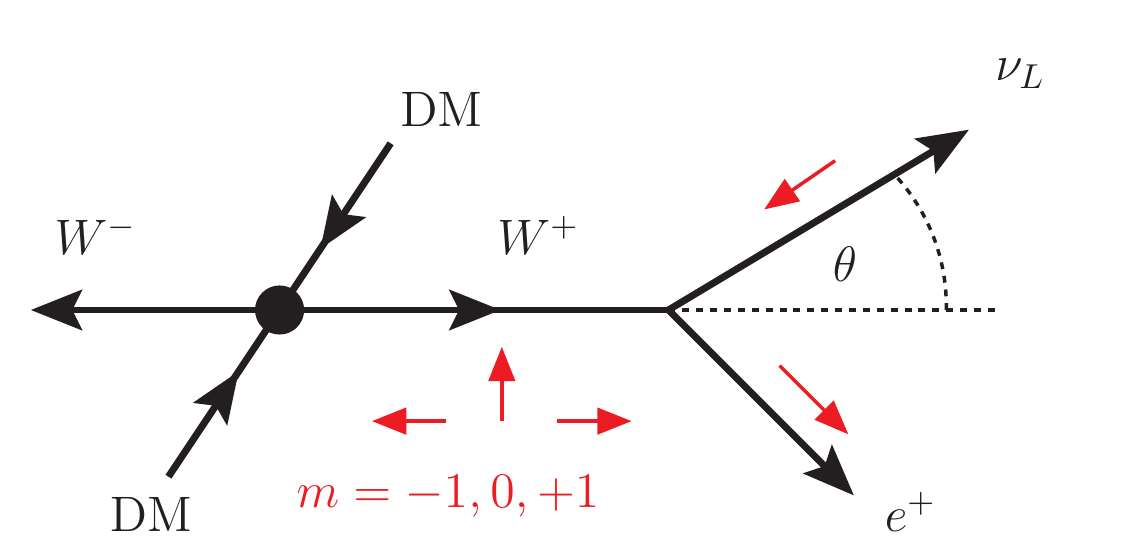}
\caption{Illustration of $\text{DM DM} \to W^+ W^-$ followed by $W^+\to e^+\nu$. The red arrows denote the polarization of particles along their direction of motion; the $W^+$ has three possible polarizations, $m=-1$, $0$, $+1$. 
}
\label{fig:polarizedWdecay}
\end{figure}

Sticking with the $\text{DM DM} \to W^+ W^-$ example, let us derive the resulting neutrino spectrum from the subsequent decay $W^+\to \ell^+\nu$ of the boosted $W$. 
If a $W^+$ boson is emitted with polarization $m \in \{-1,0,+1\}$ and energy $E_W = M_\text{DM}> M_W$
in the process $\text{DM DM} \to W^+ W^-$, it decays into $\nu e^+$ with a rate $\gamma^{-1}\Gamma_W \Br (\nu e^+)$, time dilated in the annihilation frame by the Lorentz factor $\gamma = E_W/M_W$, but independent of $m$. However, depending on the polarization $m$, the neutrino will be emitted at a preferred angle $\theta$ relative to the $W$-emission/polarization axis (see Fig.~\ref{fig:polarizedWdecay} and Ref.~\cite{Jungman:1994jr}). This translates into a different energy in the DM center-of-mass frame,
\begin{align}
E_\nu = \frac{M_W^2/2}{E_W-\sqrt{E_W^2-M_W^2} \cos\theta}  \,.
\end{align}
As expected, the largest neutrino energy $E_\nu^+$ comes from the smallest $\theta$, i.e.~those neutrinos emitted in the same direction as the~$W$ boson. The kinematic endpoints for $E_\nu$ are thus $E_\nu^\pm = (E_W\pm \sqrt{E_W^2-M_W^2})/2$. In between these endpoints, i.e.~inside this kinematic \emph{box}, the spectral shape depends on the polarization $m$ of the intermediate $W^+$ boson,
\begin{align}
\frac{\dd N_{\nu,m}}{\dd E_\nu} = \frac{f_m (E_\nu/M_\text{DM})}{M_\text{DM}} \,,
\label{eq:dNdEnu}
\end{align}
with (setting $r_W \equiv M_W/M_\text{DM}$)
\begin{align}
\begin{split}
f_0 (x) = & \frac32\, \frac{4x -4 x^2-r_W^2}{ \left(1-r_W^2\right)^{3/2}} \\
&\quad\times \Theta \left(x - \tfrac{E^-_\nu}{M_\text{DM}} \right)\Theta \left(\tfrac{E^+_\nu}{M_\text{DM}} -x\right) , \label{eq:fL}
\end{split}\\
\begin{split}
f_{\pm 1} (x) =&\frac34 \, \frac{ 2-4 x+4 x^2-r_W^2\pm (2-4 x) \sqrt{1-r_W^2}}{\left(1-r_W^2\right)^{3/2}}\\
&\quad\times \Theta \left(x - \tfrac{E^-_\nu}{M_\text{DM}} \right)\Theta \left(\tfrac{E^+_\nu}{M_\text{DM}} -x\right) ,
\end{split}
  \label{eq:fT}
\end{align}
$\Theta(x)$ being the Heaviside step function.
The normalization is such that $N_\nu$ counts the number of neutrinos emitted per $W^+\to \nu e^+$ decay,
\begin{align}
N_\nu =\int_{E_\nu^-}^{E_\nu^+}\dd E_\nu\, \frac{\dd N_{\nu,m}}{\dd E_\nu} =1\,.
\end{align}
For an unpolarized $W$ we recover a flat-box spectrum,
\begin{align}
\frac13 \sum_m f_m (x) = \frac{ \Theta \left(x - \frac{E^-_\nu}{M_\text{DM}} \right)\Theta \left(\frac{E^+_\nu}{M_\text{DM}} -x\right)  }{\sqrt{1-r_W^2}}\,.
\end{align}
The anti-neutrino spectra from $W^-\to \bar\nu e^-$ can be obtained from Eq.~\eqref{eq:dNdEnu} via $N_{\bar\nu,m} = N_{\nu,-m}$,\footnote{From our polarized spectra one can also calculate the second moments $\langle E_\nu^2 \rangle$, which gives $\langle E_\nu^2 \rangle = (E_W/2)^2 (1+\beta^2/3)$ for unpolarized decays, $(E_W/2)^2(1+\beta^2/5)$ for longitudinal, and $(E_W/2)^2 (1+2\beta^2/5\pm \beta)$ for the transversal polarizations, where $\beta = \sqrt{1-M_W^2/E_W^2}$ is the velocity of the $W$ boson in the DM rest frame. This agrees with Refs.~\cite{Kamionkowski:1991nj,Jungman:1994jr} (up to a typo).}
and the discussion of $Z\to\bar\nu\nu$ is completely analogous.

\begin{figure}[t]
\includegraphics[width=0.45\textwidth]{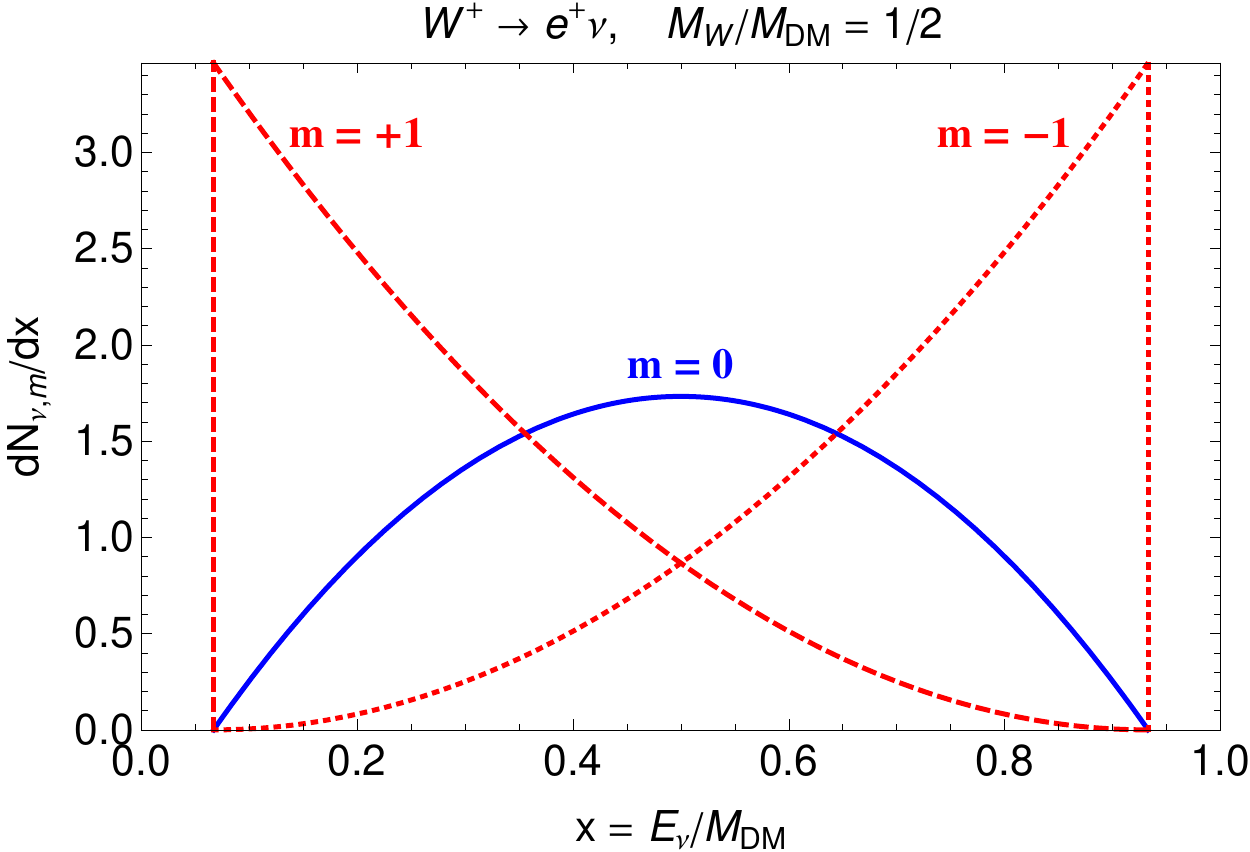}
\caption{
Neutrino spectra from Eq.~\eqref{eq:dNdEnu} for longitudinally polarized $W$ (blue) and transverse $W$ (red).}
\label{fig:spectra}
\end{figure}

We show the neutrino spectra for the three different $W$ polarizations in Fig.~\ref{fig:spectra}.
An unpolarized $W$ gives a flat box-shaped spectrum, similar to the one studied in Ref.~\cite{Ibarra:2012dw} for decaying scalar particles produced in DM annihilations. In contrast, when the gauge boson is polarized, the flat box is deformed into different shapes which are determined by the initial polarization.
A longitudinally polarized $W$ leads to a concave shape around the center of the kinematic box; the two transverse $W$ modes add up to a convex spectrum.
Note that unpolarized DM typically gives $\text{Br}_{m} = \text{Br}_{-m}$, so we effectively have to average over the $m=\pm 1$ spectra in most cases (see the examples above). The crucial point is that even for \emph{unpolarized} DM, we generically end up with different rates for longitudinal vs.~transversal $W$ bosons, courtesy of the GBET. 

\begin{figure*}[t]
\includegraphics[trim=0cm 0cm 0cm 0cm,clip,height=0.33\textwidth]{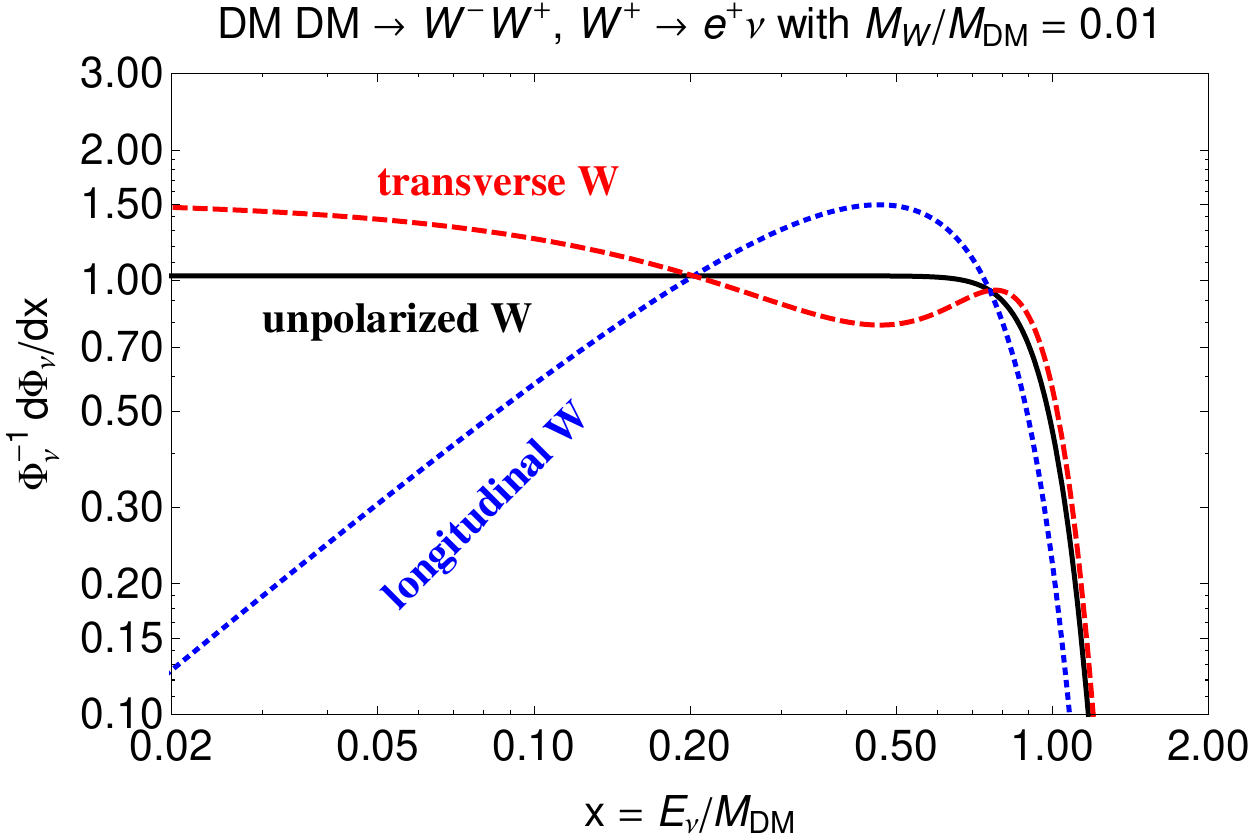}
\includegraphics[trim=0.7cm 0cm 0cm 0cm,clip,height=0.33\textwidth]{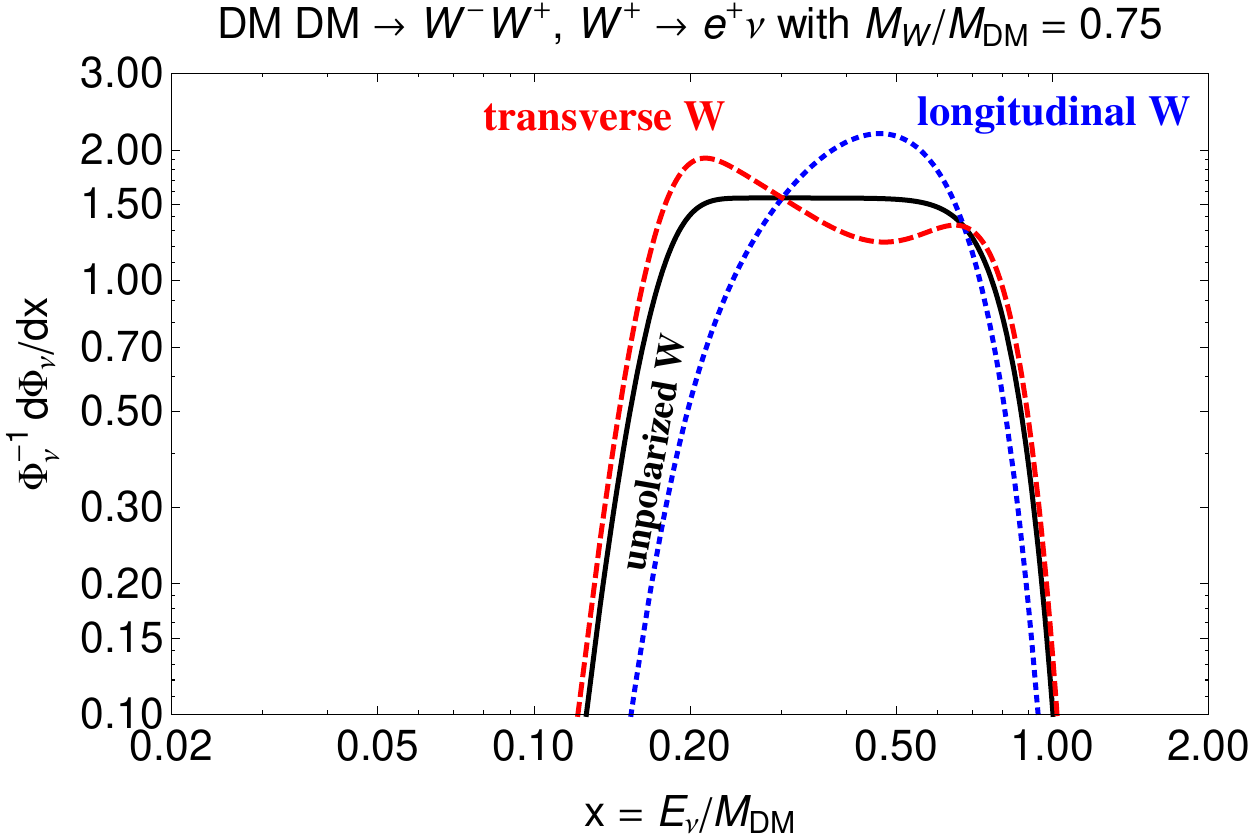}
\caption{
Neutrino differential flux from $\text{DM DM} \to W^+ W^-$ followed by $W^+\to e^+\nu$ for different $W^+$ polarizations. The spectra were folded with a Gaussian to emulate an energy resolution of the detector of $15\%$. 
}
\label{fig:DMDMtoWWneutrinospectrum}
\end{figure*}

With the production branching ratios of the different $W$ polarizations and their associated neutrino spectra, we can calculate the total neutrino flux from $\text{DM DM} \to W^- (W^+\to e^+ \nu)$ as
\begin{align}
\frac{\dd \Phi_\nu}{\dd E_\nu} &= \Phi_\nu
\sum_m \Br_m \frac{\dd N_{\nu,m}}{\dd E_\nu}  \,,& \Phi_\nu= \frac{(\sigma v) }{8\pi M_\text{DM}^2} \bar{J}_\text{ann} \,,
\label{eq:neutrinoflux}
\end{align}
where $(\sigma v)\equiv \sigma v (\text{DM DM} \to  W^- W^+)\times\Br (W^+\to e^+ \nu)$ is the total cross section for the process and the astrophysical factor $\bar{J}_\text{ann}$ is defined as
\begin{align}
\bar{J}_\text{ann} \equiv \frac{1}{\Delta \Omega} \int_{\Delta \Omega} \dd \Omega \int_{\text{l.o.s.}} \dd s\, \rho_\text{DM}^2\,,
\label{eq:Jfactor}
\end{align}
where $\rho_\text{DM}$ is the DM density and $s$ the distance along the line of sight (l.o.s.).
 For Dirac DM, the flux is a factor $1/2$ smaller compared to the Majorana DM assumed above.
Since the energy spectrum is given by $\sum_m \Br_m \frac{\dd N_{\nu,m}}{\dd E_\nu}$, if either the production or the decay is unpolarized, the polynomial feature disappears and we obtain a flat-box shape, as in the case of a scalar mediator~\cite{Ibarra:2012dw}.
We show this quantity in Fig.~\ref{fig:DMDMtoWWneutrinospectrum} for the three cases of transverse $W^+$ (e.g.~wino DM, Eq.~\eqref{eq:Pwino}), longitudinal $W^+$ (e.g.~singlet scalar DM, Eq.~\eqref{eq:Psinglet}), and unpolarized $W^+$ (the familiar flat-box spectrum, achievable e.g.~by tuning in the Inert Doublet model, Eq.~\eqref{eq:PIDM}).
Note the log--log axes compared to Fig.~\ref{fig:spectra}.
To estimate a somewhat realistic flux, we have folded the spectrum with a Gaussian distribution to model an energy resolution of $15\%$. (Because the resolution increases with the energy, the height of those spectral features -- originally equal according to Fig.~\ref{fig:spectra} -- decreases for larger energies.) This partially washes out the characteristic $m=-1$ spike of Fig.~\ref{fig:spectra} into a generic spectral feature. As a result, experimental limits on the flux will not depend strongly on the polarization, and hence on the $SU(2)_L$ quantum numbers of DM. Nonetheless, an improvement on the experimental resolution of neutrino detectors can overcome this difficulty and allow to distinguish the various spectral shapes.

In most cases $\Br_m = \Br_{-m}$ and thus the spectrum for anti-neutrinos from the $W^-\to e^- \bar\nu$ will be identical to the neutrino spectrum. The same is true for $Z\to\nu\bar\nu$ up to factors of two if the $Z$ is produced in pairs.
This concludes our example and shows that the polarization of intermediate particles in DM annihilation generically plays a role for the resulting spectra.

\section{Generalized box-shaped spectra}
\label{sec:decay}

\begin{figure}[b]
\includegraphics[width=0.40\textwidth]{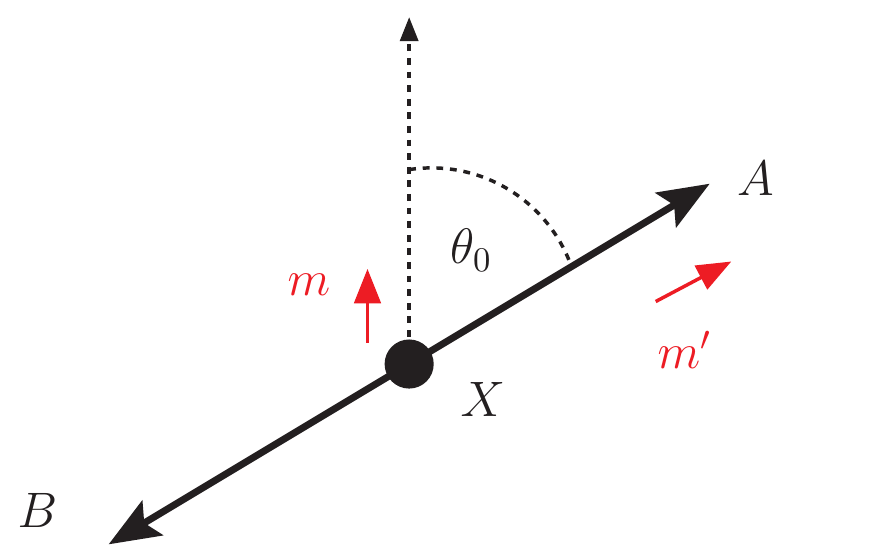}
\caption{ Decay of the intermediate particle $X$ in its rest frame. The decay products $A$ and $B$ form an angle $\theta_0$ with respect to the direction of motion of $X$ in the annihilation frame (i.e.~the boost direction). The red arrows sketch the total angular momentum  of the initial and final states.  
}
\label{fig:restframeX}
\end{figure}

We have seen that the polarization of the $W$ bosons emitted from DM annihilations have remarkable consequences for the spectral shape of their decay products. In this section we extend this result to other particles with arbitrary polarizations.

Suppose that DM annihilates into two particles, one of which we call $X$ and whose spin is $S$. Moreover, assume that $X$ in turn decays into two particles $A$ and $B$ with negligible mass, so that $\text{DM DM}\to Y (X\to AB)$.  The key aspect to have in mind is that the polarization of $X$ -- which is just the angular momentum along its direction of motion -- determines the angular distribution of its decay products in its rest frame. There, particles $A$ and $B$ move back-to-back forming an angle $\theta_0$ with respect to the original direction of motion of $X$ (see Fig.~\ref{fig:restframeX}). Each angle $\theta_0$ in the rest frame corresponds to a fixed energy $E_A$ for $A$ in the frame where DM annihilates,
\begin{align}
E_A = \frac{1}{2} \left(E_X+\sqrt{E_X^2-M_X^2} \cos\theta_0\right) ,
\label{eq:EA}
\end{align}
so we can calculate the energy spectrum of $A$  once the angular distribution in the rest frame is known.   In order to calculate the latter,  suppose that $m$ is the angular momentum of $X$ along the boost direction and that $m'$ is the total angular momentum along the direction of motion of the decay products.  Notice that since we are sitting at the rest frame of $X$, its total angular momentum is simply its spin. Moreover, since the orbital angular momentum of the decay products along the direction of their motion vanishes, $m'$ is just the helicity difference of $A$ and $B$. 
Then, the initial and final states are described by  $|m,S\rangle$ and  $|m',S\rangle^{\theta_0} = R(\theta_0)^{T}|m',S\rangle$, respectively. Here,  $R(\theta_0)$ is a rotation operator taking  the direction of motion of the particles $A$ and $B$ into the boost direction  as shown in Fig.~\ref{fig:restframeX}. The probability amplitude for the decay process must then be proportional to 
\begin{align}
{}^{\theta_0} \langle m', S | m, S\rangle = \langle m',S| R(\theta_0) | m, S\rangle \equiv  d^S_{m'm} (\theta_0)\,, 
\end{align}
where we introduce the Wigner $d$-functions~\cite{wigner1931gruppentheorie}. As is clear from the discussion, these are just representation matrices for ordinary rotations in the spin space with total angular momentum $S$. 

In order to find the angular distribution for a given polarization $m$ of the parent particle,  we just have to add the contribution of each helicity combination of the final state. This means that the angular distribution in the rest frame of $X$ is given by
\begin{align}
\frac{\dd N_{A,m}}{\dd \cos\theta_0} \propto \sum_{m'} C_{m'}|d^S_{m'm}(\theta_0)|^2\,,
\label{eq:polarized_decayTheta00}
\end{align} 
where $C_{m'}$ are model-dependent positive coefficients that determine how the different polarizations couple to the particles $A$ and $B$. Rotational invariance in the rest frame dictates  that these coefficients are independent of $m$. Also, without loss of generality, we normalize them to one, $\sum_{m'} C_{m'} =1$. Using the fact that $\int_0^\pi \dd\theta \sin\theta \,|d^S_{mm'}(\theta)|^2 = 2/(2S+1)$, we find  
\begin{align}
\frac{\dd N_{A,m}}{\dd \cos\theta_0} = n \left(\frac{2S+1}{2}\right)\sum_{m'} C_{m'}|d^S_{m'm}(\theta_0)|^2\,,
\label{eq:polarized_decayTheta0}
\end{align} 
where $n$ is the number of $A$ particles produced in the decay of $X$ ($n=1$ or $n=2$ for $B=A$).

We must now boost this result from the rest-frame to the DM annihilation frame. 
The polarization of $X$ does not change by boosting from one frame to another as long as we use the direction of motion of the particle $X$ as our quantization axis. The underlying reason for this is the fact that the component of the orbital angular momentum along the direction of motion is zero in both frames. Using this and Eq.~\eqref{eq:EA} we find 
\begin{align}
\begin{split}
\frac{\dd N_{A,m}}{\dd E_A}& = \frac{n}{M_\text{DM}}f^S_{m} \left(\frac{E_A}{M_\text{DM}},\frac{E_X}{M_\text{DM}}\right) ,
\label{eq:dNdEA}
\end{split}
\end{align}
where
\begin{align}
\begin{split}
f^S_m (x,y) &=  \frac{(2S+1)}{\sqrt{y^2-r_X^2} } \Theta \left(x - x^-(y)\right)\Theta \left(x^+(y) -x\right) \\ 
& \times\sum_{m'} C_{m'}\left|d^S_{m'm}\left(\arccos\left(\frac{2x-y}{\sqrt{y^2-r_X^2}}\right)\right)\right|^2 .
\end{split}
\label{eq:fsm}
\end{align} 
with $r_X=M_X/M_\text{DM}$ and $x^\pm(y) =(y\pm \sqrt{y^2-r_X^2})/2 $.  Notice that for a given process, $y = E_X/M_\text{DM}$ is fixed by kinematics. For instance, the process $\text{DM} \, \text{DM}\to X\overline{X}$ gives $y=1$.  The previous equation can now be used along with a generalization of Eq.~\eqref{eq:neutrinoflux} to calculate the differential flux for $A$ particles 
\begin{align}
(\Phi_A)^{-1}\frac{\dd \Phi_A}{\dd E_A}&=  \frac{1}{n} \sum_m \Br_m\frac{\dd N_{A,m}}{\dd E_A}  \label{eq:Aflux}\\
&= \frac{1}{M_\text{DM}}
\sum_m \Br_m f^S_{m} \left(\frac{E_A}{M_\text{DM}},\frac{E_X}{M_\text{DM}}\right) .\nonumber
\end{align}
Note that although we derived our formulae for DM annihilations, they can also be applied for decays.
Before discussing some general properties of Eq.~\eqref{eq:Aflux}, we first consider some particular examples for the sake of illustration.

\subsection{Decaying scalars}  

Taking $X$ to be a scalar particle, i.e.~with spin $S=0$, severely simplifies Eq.~\eqref{eq:fsm}.
In this case, the Wigner $d$-functions are trivially equal to one, the spectrum of particles $A$ is flat for energies allowed by kinematics and zero otherwise. We thus recover the flat box-shaped spectra that was first discussed in Ref.~\cite{Ibarra:2012dw} in the context of gamma-ray spectral features, taking place when a scalar produced in DM annihilations subsequently decays into photons. Such spectra were also previously discussed in the context of cosmic rays~\cite{stecker1971cosmic}. Notice that from our general arguments, this conclusion is not only true for photons but also for neutrinos or any other light particle.

As an example of a scalar mediator within the SM, let us consider the case of DM annihilating into the SM scalar $h$. This will generate flat gamma-ray boxes from the decay $h\to \gamma\gamma$. Another example consists of axion-like particles produced in DM annihilations which then decay into two photons (see e.g.~Ref.~\cite{Garcia-Cely:2013wda}).

\subsection{Decaying fermions} 

The next-to-simplest possibility is the case where the decaying particle $X$ has spin $S=\oh$. Here, there are two independent Wigner $d$-functions, $d^{\oh}_{\pm\oh\,\oh}(\theta_0) = \pm\sqrt{(1\pm\cos\theta_0)/2}$, so
\begin{align}
\begin{split}
f^{\oh}_{\pm \oh}(x,y)&=  \frac{1}{\sqrt{y^2-r_X^2}} \Theta \left(x - x^-(y)\right)\Theta \left(x^+(y) -x\right)  \\ 
&\quad\times \left( 1 \pm \left(C_{\oh}-C_{-\oh}\right) \left(\frac{2x-y}{\sqrt{y^2-r_X^2}} \right) \right) ,
\end{split}
\label{eq:polarized_decay12}
\end{align} 
linear in $x$.
Therefore, the decay of a spin-$\oh$ particle leads to spectra given by a straight line, or  triangle-shaped spectra. 
This scenario has been discussed in Ref.~\cite{Ibarra:2016fco} in the context of spectral features for asymmetric DM. The slope of the triangle box is determined by how differently the final states with helicity $\pm\oh$ couple. Clearly, if there is no physical mechanism that distinguishes both states, one has $C_{\oh}=C_{-\oh}$ and the flat box-shaped spectrum is recovered.

As an example, suppose that (asymmetric) DM is charged with positive $\tau$-lepton number and that it decays into a $\tau^-$ along with some other charged particle. The former decays 10.83\%  of the time into $\pi^- \nu_\tau$~\cite{Agashe:2014kda}. This leads to a spectrum of tau neutrinos determined by Eqs.~\eqref{eq:Aflux} and \eqref{eq:polarized_decay12} with $C_{-\oh}=  1$ and $C_{\oh}= 0$, because the final state (left-handed neutrino plus scalar meson) can only have one helicity.   

Likewise, for an intermediate particle with spin $\tth$, the spectrum is a kinematic box with a shape given by a cubic function of the energy. All the possible spectral shapes, as given by the corresponding angular distributions, are reported in Fig.~7 of Ref.~\cite{Christensen:2013aua}.

\subsection{Decaying vectors}  

Let $X$ be a massive vector boson, similar to the $W$ of our example in Sec.~\ref{sec:Wbosondecay}.
In this case, the diphoton final state does not exist because of the Landau--Yang theorem~\cite{Landau,Yang:1950rg}, so we focus on the case of gauge bosons decaying into two massless fermions with helicity $\pm\oh$.  There, we only have the final states with $m' = \pm 1$  because those with  $m'=0$ would require a mass insertion (see below). This immediately leads to $C_0 =0$. For each polarization of the intermediate particle $X$, there are only two relevant Wigner $d$-functions. For a longitudinal polarized boson, one has $m=0$ and $d^1_{\pm1\,0}(\theta_0)= \pm \sin\theta_0/\sqrt{2}$. Therefore,
\begin{align}
\begin{split}
f^1_0 (x,y) &=  \Theta \left(x - x^-(y)\right)\Theta \left(x^+(y) -x\right)\\  
&\quad \times \frac32 \frac{4x y-4x^2-r_X^2}{(y^2-r_X^2)^{3/2} }  
 \,,
\end{split}
\label{eq:polarized_decay1L}
\end{align} 
which is Eq.~\eqref{eq:fL} for $y =E_X/M_\text{DM} =1$. Notice that the dependence on the couplings $C_{\pm1}$ drops out because the corresponding squared Wigner $d$-functions are equal. For the transverse mode the spin component is $m = \pm 1$,  we then have $d^1_{m'\pm1} (\theta_0) = (1\pm m'\cos\theta_0)/2$ and the corresponding spectrum is given by
\begin{align}
\begin{split}
&f^1_{\pm1}(x,y)=  \frac{3  \Theta \left(x - x^-(y)\right)\Theta \left(x^+(y) -x\right)}{4(y^2-r_X^2)^{3/2} } \\
&\quad\times \left[2y^2 -4xy + 4x^2  -r_X^2\right.  \\
&\qquad\left. \pm 2\left(C_{+1}-C_{-1}\right)(2x-y)\sqrt{y^2-r_X^2}\right] .
\end{split}
\label{eq:polarized_decay1T}
\end{align} 
This agrees with Eq.~\eqref{eq:fT} for $y =E_X/M_\text{DM} =1$ if we take $C_{-1}= 1$ and $C_{+1} =0 $. This is a consequence of the fact that the $W$ boson only couples to left-handed fermions (and right-handed antifermions). In fact, for a vector boson $X$ coupled to the fermionic current $\bar{B}\gamma^\mu (g_L P_L+g_R P_R) A$, we obtain to leading order in $M_{A,B}/M_X$, 
\begin{align}
\begin{split}
C_0 & \simeq \frac{M_A^2+M_B^2}{2 M_X^2}+ \frac{2\, g_L g_R }{g_L^2+g_R^2} \frac{M_A M_B}{M_X^2} \,,\\
  C_{-1} &\simeq \frac{g_L^2(1-C_0)}{g_L^2+g_R^2} - \frac{2\, g_L g_R (g_L^2-g_R^2) }{(g_L^2+g_R^2)^2} \frac{M_A M_B}{M_X^2}   \,, \\
C_{+1} &\simeq \frac{g_R^2(1-C_0)}{g_L^2+g_R^2} + \frac{2\, g_L g_R (g_L^2-g_R^2) }{(g_L^2+g_R^2)^2} \frac{M_A M_B}{M_X^2}  \,.
\end{split}
\label{eq:VtoAB}
\end{align}
From these expressions we explicitly see that the state with $m'=0$ is mass-suppressed. Also, note that when parity is conserved, $g_L = g_R$ and thus $ C_{-1} = C_{+1}$.

Having dismissed the diphoton final state due to the Landau--Yang theorem, let us mention that anomalous processes such as $Z'\to Z\gamma$ can occur via Chern--Simons couplings. If the $Z'$ is coupled to e.g.~fermionic DM, this can give rise to monochromatic photons via $s$-channel $Z'$ exchange~\cite{Dudas:2009uq}. The same model will lead to quadratic gamma-ray spectra within a box from the annihilation channel $\text{DM DM}\to Z' Z'$, followed by $Z'\to Z\gamma$ according to our formulae. The corresponding coefficients are
\begin{align}
  C_{\pm1} =&\frac{M_{Z'}^2}{M_{Z}^2+2 M_{Z'}^2} \,, & C_0 = \frac{M_{Z}^2}{M_{Z}^2+2 M_{Z'}^2} \,.
	\label{eq:chern-simons}
\end{align}
As in the previous case, we find a mass suppression for the $m'=0$ state when $M_Z \ll M_{Z'}$.

\subsection{Decaying tensors}

\begin{figure*}[t]
\includegraphics[height=0.3\textwidth]{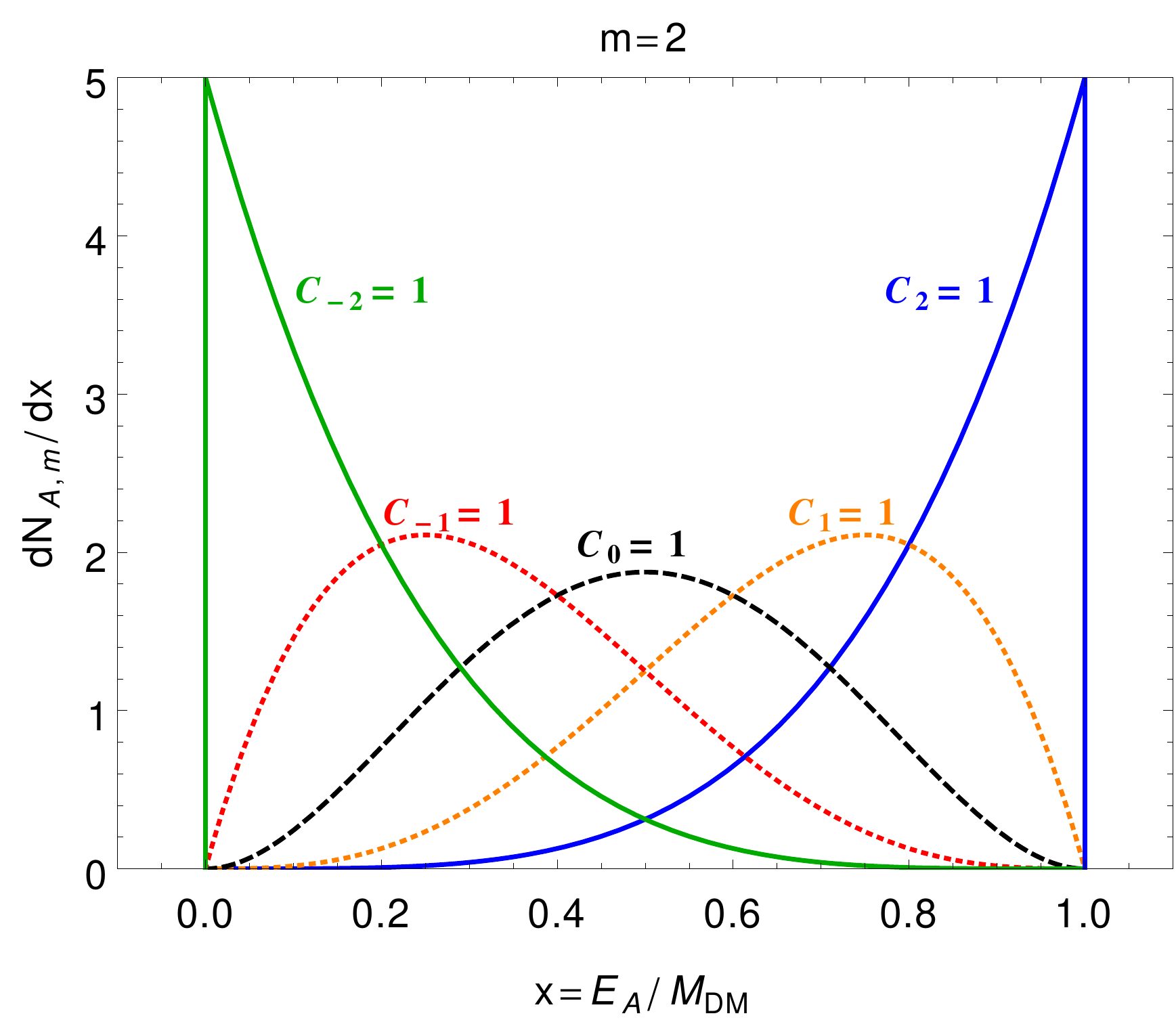}
\includegraphics[trim=1.7cm 0cm 0cm 0cm,clip,height=0.3\textwidth]{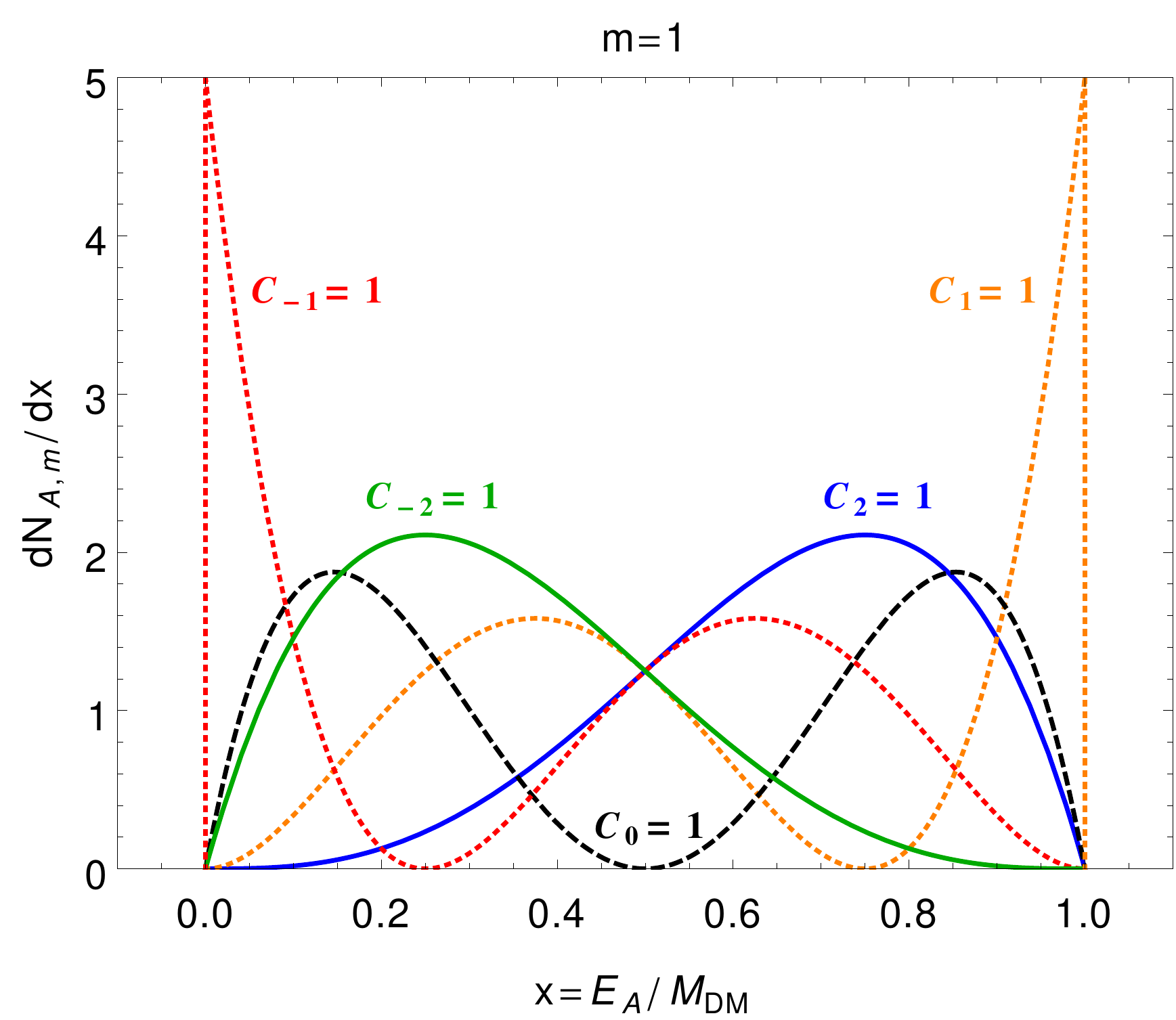}
\includegraphics[trim=1.7cm 0cm 0cm 0cm,clip,height=0.3\textwidth]{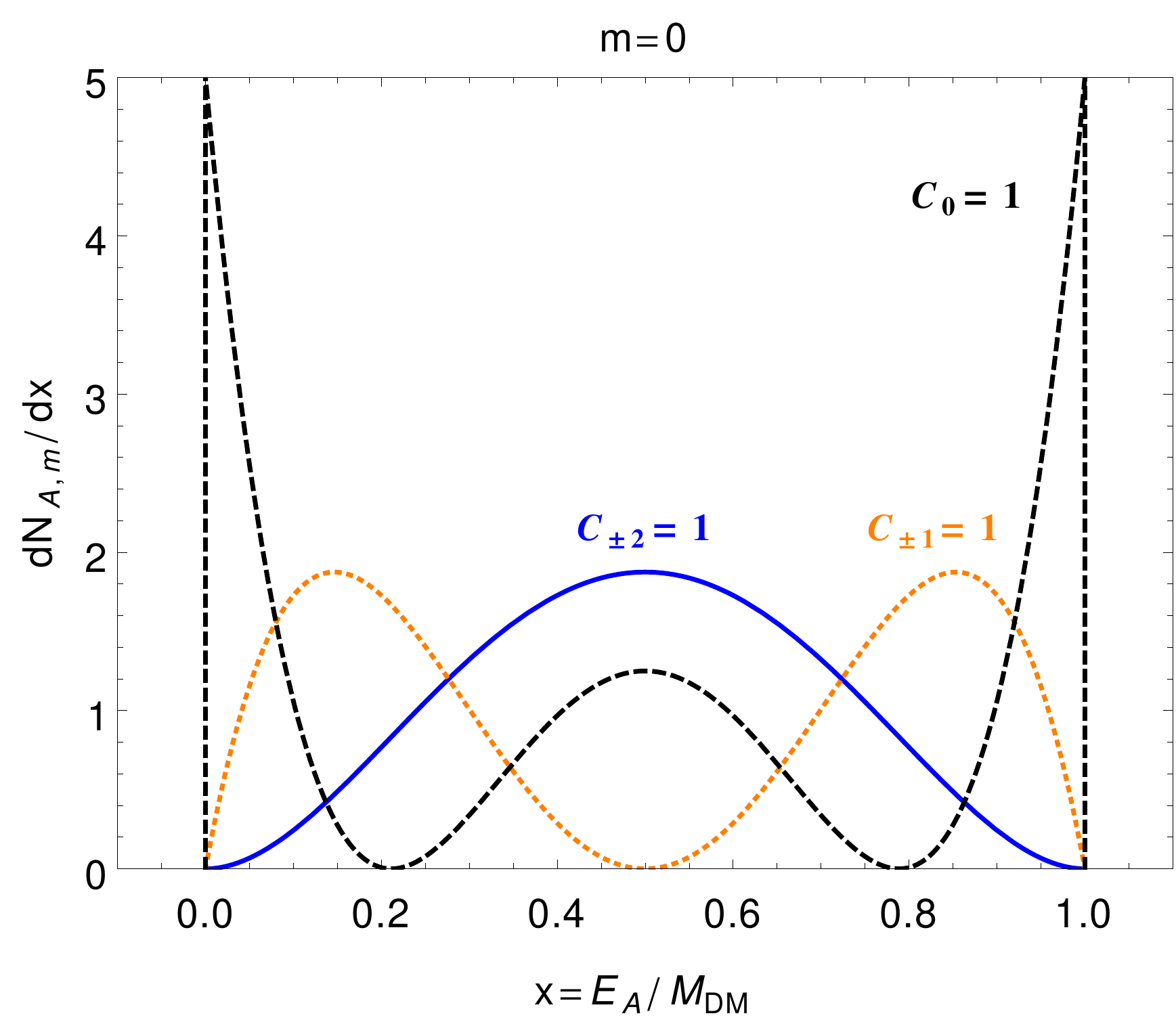}
\caption{Energy spectra of $A$ given by Eq.~\eqref{eq:dNdEA} from a decaying spin-2 particle $T\to AB$ arising from DM annihilations $\text{DM DM}\to Y T$ with polarizations $m=2$ (left), $m=1$ (center), and $m=0$ (right), assuming a fixed helicity $m'$ in the final state, i.e. $C_{m'}=1$. 
Any other spectrum is a superposition of the ones shown here.  
The spin-2 particle is assumed to be much lighter than the DM (i.e.~$r_X\ll1$) and its energy equal to $M_\text{DM}$ (i.e.~$y=1$). A different choice of $r_X$ will only shrink the box.
}
\label{fig:BoxesSpin2}
\end{figure*}

For spin-$2$ particles $T$ produced by DM, the spectra of $A$ from $T\to AB$ are quartic in $E_A$.  In Fig.~\ref{fig:BoxesSpin2}, we depict the ones associated to $m=2$ (left plot), $m=1$ (central plot), and $m=0$ (right plot) when each of the coefficients $C_{m'}$ entering in Eq.~\eqref{eq:fsm} is set to one. The remaining spectra can be obtained from the latter by reflection around the central point (see discussion below). Notice that according to Eq.~\eqref{eq:fsm}, any other spectrum is merely a superposition of the functions shown in Fig.~\ref{fig:BoxesSpin2}.

As a well-motivated example, we consider a massive spin-$2$ particle $T$ that is coupled to the energy--momentum tensor of its decay products (see Sec.~\ref{sec:production}). Using the Feynman rules from Ref.~\cite{Han:1998sg}, one can calculate the corresponding coefficients $C_{m'}$. We give some of them in Tab.~\ref{table:TensorC}, including the most relevant decays into photons and neutrinos (be it purely left-handed or massive). 
For the massive gauge bosons, $T\to ZZ$, $W^+W^-$, we find $C_0 =\frac{1}{13}$ in contrast to the diphoton case, $T\to\gamma\gamma$, for which $C_0 = 0$. This contribution can be traced back to the Goldstone modes associated to the massive part of the gauge bosons, and is closely related to the mismatch of the total decay rates $\Gamma (T\to \gamma\gamma)/\Gamma (T\to ZZ) \xrightarrow{M_Z\to 0} 12/13$~\cite{Han:1998sg}.

\begin{table}[t]
  \begin{center}
    \begin{tabular}{c|c|c|c|c|c}
final state $AB$ & $C_{-2}$ &$C_{-1}$ &$C_{0}$ & $C_{1}$ &$C_{2}$ \\\hline
$\gamma\gamma$ & $\frac{1}{2}$ & 0 & 0 & 0 & $\frac{1}{2}$\\
$ZZ$ &  $\frac{6}{13}$  & 0 &$\frac{1}{13}$ &0 & $\frac{6}{13}$\\
$W^+W^-$ &  $\frac{6}{13}$  & 0 &$\frac{1}{13}$ &0 & $\frac{6}{13}$\\
$hh$ & 0 & 0 &1 &0 &0 \\
$\nu_L\overline{\nu_L}$ &  0 & 1 & 0 &0 & 0\\
$\nu_R\overline{\nu_R}$ &  0 & 0 & 0 &1 & 0\\
$\nu\overline{\nu}$ (Dirac or Majorana) &  0 & $\frac{1}{2}$ & 0 & $\frac{1}{2}$ & 0
    \end{tabular}
  \end{center}
\caption{$C_{m'}$ coefficients in Eq.~\eqref{eq:fsm} leading to the energy spectrum of decaying spin-2 particle $T\to AB$ coupled to the energy--momentum tensor. In the case when there are particles and antiparticles, our convention is that $A$ corresponds to the \emph{particle}. For the spectrum of the \emph{antiparticle}, the coefficients are given by $C_{m'}(B) = C_{-m'}(A)$.}
  \label{table:TensorC}
\end{table}

We will discuss the spin-2 tensor in more detail in Sec.~\ref{sec:production} when we give examples for its production. Before that, let us conclude this section by discussing some general features of Eqs.~\eqref{eq:fsm}--\eqref{eq:Aflux}.

\subsection{General properties}

For any intermediate unstable spin-$S$ particle $X$ of an annihilation $\text{DM DM}\to Y(X\to AB)$, the spectrum of $A$ particles has the following properties, according to Eq.~\eqref{eq:fsm}:
\begin{itemize}
\item It is a kinematic box with endpoints $x^-(y)$ and $x^+(y)$, centered at $x = y/2$,  and with an overall shape given by a superposition of the polynomials of order $2S$,
\begin{align}
\begin{split}
&|d^S_{m'm} (\arccos t)|^2 =(S+m')!(S-m')!(S+m)!(S-m)!\\
&\left(\sum_{s}\frac{(-1)^{m'-m+s} \left(1+ t\right)^{\frac{2S+m-m'-2s}{2}}\left(1-t\right)^{\frac{m'-m+2s}{2}}}{2^Ss!(S+m-s)!(m'-m+s)!(S-m'-s)!} \right)^2 \!,
\end{split}
\end{align}
where the variable $s$ runs over all possible integers such that all the factorials involved are non-negative. 
The polynomial structure of certain spectra has been discussed in Ref.~\cite{Tang:2015meg} using other methods. Similarly, polynomial kinematic distributions have been studied in Ref.~\cite{Wang:2006hk} in the context of colliders.
\item From the identity $d^S_{m',-m}(\theta_0)=(-1)^{S+m'}d^S_{m',m}(\pi-\theta_0)$, it follows that intermediate states $X$ with opposite polarizations give rise to spectra that are specular images of each other which respect to the central point $x = y/2$. Likewise, the spectra associated to $m=0$ are symmetric around the central point. All of this is illustrated in Figs.~\ref{fig:spectra} and~\ref{fig:BoxesSpin2}.
\item  The role of $B$ and $A$ can be interchanged by means of a reflection along their direction of motion (see Fig.~\ref{fig:restframeX}). Hence, the flux for $B$ is obtained by taking $C_{m'}(B) = C_{-m'}(A)$ in Eq.~\eqref{eq:fsm}.  Equivalently,  $N_{B,m} = N_{A,-m}$ due to $|d^S_{m',-m}|^2 = |d^S_{-m',m}|^2$ and therefore
\begin{align}
\qquad \frac{\dd \Phi_B}{\dd E_B} &= \frac{\Phi_B}{M_\text{DM}}
\sum_m \Br_m f^S_{-m} \left(\frac{E_B}{M_\text{DM}},\frac{E_X}{M_\text{DM}}\right) .
\label{eq:Bflux}
\end{align}
The flux of $B$ is thus a specular image of the flux of $A$ with respect to the central point. In particular, for identical particles in the final state, $B=A$, the spectrum is symmetric.
\item Suppose that  all the coefficients $C_{m'}$ are equal. Then $C_{m'} = 1/(2S+1)$ by normalization and from $\sum_{m'} |d_{m'm}(\theta_0)|^2 = 1$ we obtain
\begin{equation}
f^S_m (x,y) =  \frac{\Theta \left(x - x^-(y)\right)\Theta \left(x^+(y) -x\right) }{\sqrt{y^2-r_X^2} } \, .
\label{eq:Cequals}
\end{equation} 
Hence, a non-flat spectrum can only arise  if the polarizations of the final states behave in a different manner. For instance, this is the case of spin-$\oh$ decaying fermions, in which the ``triangle-shaped'' spectrum only arises if the left- and right-handed decay products couple differently~\cite{Ibarra:2016fco}. Similarly, in gauge boson decays, we have seen that final states with helicity $m'= 0$ are typically negligible since they are mass-suppressed (see Eqs.~\eqref{eq:VtoAB}--\eqref{eq:chern-simons}).
\item If the intermediate particle $X$ is \emph{produced} unpolarized,  the energy spectrum is proportional to $\sum_m  f^S_m$, i.e.~each polarization contributes with the same weighting factor. From  $\sum_{m'} |d_{m'm}(\theta_0)|^2 = 1$, we again obtain a flat spectrum. Thus,  even if there exists a mechanism that distinguishes the different helicities of the final state (such as chiral couplings or mass-suppression), in order to have non-flat spectra, one also needs another mechanism differentiating the spins of the intermediate decaying state, otherwise the total rate erases the different spectral features into a flat-box shape. In the case of the particles of spin-$\oh$ this can be achieved by means of an asymmetry in the DM sector. In the case of massive gauge bosons, as discussed above, this happens automatically at high energies since their longitudinal and transverse components are identified with two completely different fields, the Goldstone bosons and the massless gauge fields, respectively. As we will see in the following section, this is also the case of spin-2 particles produced in DM annihilations.
\end{itemize}

Before finishing this section, we would like to comment on an algorithm to calculate the coefficients $C_{m'}$ in an arbitrary model. Since $d^S_{m'm}(0) = \delta_{m'm}$, according to Eq.~\eqref{eq:polarized_decayTheta0}, we have
\begin{align}
 C_{m'}=  \frac{2}{(2S+1)\,n} \frac{\dd N_{X,m'}}{\dd \cos\theta_0}\Big|_{\theta_0 =0} \,.
\label{eq:CmpAlgorithm}
\end{align} 
The coefficient $C_{m'}$ can therefore be calculated simply by considering the decay rate of the particle $X$ in its rest frame when its  spin along a given axis is $m'$ and the final two-body state is aligned with said axis. This immediately shows that $C_{-m'} =C_{m'} $ when parity is conserved.

\section{Production of spin-2 particles from DM annihilation}
\label{sec:production}

In analogy to the  case of $W$ bosons produced in DM annihilations, we will now show that spin-2 particles are also generically produced \emph{polarized}. The (Fierz--Pauli) Lagrangian of a massive spin-2 particle is given in terms of a symmetric tensor field $T_{\mu\nu}$ as~\cite{Fierz:1939ix}
\begin{align}
\L_T &= -\tfrac12 \del_\lambda T_{\mu\nu} \del^\lambda T^{\mu\nu} + \del_\mu T_{\nu\lambda} \del^\nu T^{\mu\lambda} + \tfrac12 \del_\lambda T^\alpha_\alpha \del^\lambda T^\beta_\beta \nonumber\\
&\quad- \del_\mu T^{\mu\nu} \del_\nu T^\alpha_\alpha-\tfrac12 M_T^2 (T^{\mu\nu}T_{\mu\nu} -T^\alpha_\alpha T^\beta_\beta) \nonumber\\
&\quad-\frac{1}{\Lambda} T^{\mu\nu} \mathcal{T}_{\mu\nu}^X,
\label{eq:fierz-pauli}
\end{align}
where we introduced a linear coupling to a source tensor $\mathcal{T}_{\mu\nu}^X$, to be specified later. We remain agnostic about the origin of this spin-2 particle and in particular do not necessarily assume it to be a Kaluza--Klein excitation of the graviton. For the current status on the consistent (ghost-free) Lagrangians of massive spin-2 fields beyond the linear level we refer to Refs.~\cite{Hinterbichler:2011tt,deRham:2014zqa,Schmidt-May:2015vnx}.\footnote{See also Refs.~\cite{Aoki:2016zgp,Babichev:2016hir} where the massive tensor is not just a mediator to the DM sector but is taken long-lived enough to form DM itself.}
The five physical degrees of freedom contained in $T^{\mu\nu}$ can be described by the polarization tensors
\begin{align}
\epsilon^{\mu\nu} (p,\pm 2) &= \epsilon^{\mu} (p,\pm)\epsilon^{\nu} (p,\pm)\,,\\
\epsilon^{\mu\nu} (p,\pm 1) &= \tfrac{1}{\sqrt{2}} \left[ \epsilon^{\mu} (p,\pm)\epsilon^{\nu} (p,0)+\epsilon^{\mu} (p,0)\epsilon^{\nu} (p,\pm) \right] ,\nonumber\\
\epsilon^{\mu\nu} (p,0) &= \tfrac{1}{\sqrt{6}} \left[ \epsilon^{\mu} (p,+)\epsilon^{\nu} (p,-)+\epsilon^{\mu} (p,-)\epsilon^{\nu} (p,+) \right. \nonumber\\
&\quad\left.+2 \epsilon^{\mu} (p,0)\epsilon^{\nu} (p,0) \right] \nonumber\\
       & = \tfrac{1}{\sqrt{6}} \left[ \eta^{\mu\nu} -\frac{p^\mu p^\nu}{M_T^2}+3 \epsilon^{\mu} (p,0)\epsilon^{\nu} (p,0) \right] \nonumber ,
\end{align}
built out of the familiar spin-1 polarization vectors $\epsilon^{\mu} (p,\pm) = \tfrac{1}{\sqrt{2}} (0,\pm 1,i,0)$ and $\epsilon^{\mu} (p,0) = \tfrac{1}{M_T} (p,0,0,\sqrt{M_T^2+p^2})$ where $T$ propagates in the $z$ direction with momentum $p^\mu = (\sqrt{M_T^2+p^2},0,0,p)$. 

In the limit of a boosted particle, $M_T^2 \ll p^2$, the polarization vector $\epsilon^{\mu} (p,0)$ becomes increasingly aligned with the particle momentum, i.e.~$\epsilon^{\mu} (p,0) = p^\mu/M_T + \mathcal{O}(M_T/p)$, which gives
\begin{align}
\begin{split}
\epsilon^{\mu\nu} (\pm 2) &= \epsilon^{\mu} (\pm)\epsilon^{\nu} (\pm)\,,\\
\epsilon^{\mu\nu} (\pm 1) &\simeq \tfrac{1}{\sqrt{2}M_T} \left[ p^{\nu}\epsilon^{\mu} (\pm) +p^{\mu} \epsilon^{\nu} (\pm) \right] ,\\
\epsilon^{\mu\nu} (0) &\simeq  \frac{\eta^{\mu\nu}}{\sqrt{6}}+\sqrt{\tfrac23}\,\frac{ p^{\mu}p^{\nu} }{M_T^2}\,.
\end{split}
\end{align}
This is already enough to make the following statements:
\begin{itemize}
	\item If the source is conserved, we have $\del^\mu\mathcal{T}_{\mu\nu}^X = 0$ and each tensor emission amplitude $\mathcal{M}_{\mu\nu} \bar\epsilon^{\mu\nu}(p,m)$ satisfies $\mathcal{M}_{\mu\nu} p^\mu = 0 = \mathcal{M}_{\mu\nu} p^\nu$. Hence, the emission of a boosted tensor with polarization $m =\pm 1$ is parametrically suppressed by $M_T^2/p^2$. In other words, the $m =\pm 1$ helicities decouple for $M_T \to 0$ if the source is conserved, and the Lagrangian turns into that of a massless spin-2 particle $G^{\mu\nu}$ and a massless scalar $\pi$, which couples to the trace of the source:
	\begin{align}
	\L_\mathrm{int} =-\frac{1}{\Lambda} \left(G^{\mu\nu} +\tfrac{1}{\sqrt{3}} \pi \eta^{\mu\nu}\right) \mathcal{T}_{\mu\nu}^X \,.
	\end{align}
	The ratio between the $m =0$ and $m =\pm2$ channels in this limit is model and process dependent, as we will see explicitly below.
	\item If the source is \emph{not conserved}, the dominant helicity will be $m =0$, because the rate is generically enhanced by $p^2/M_T^2$ and $p^4/M_T^4$ compared to $m =\pm 1$ and $m =\pm 2$, respectively.
\end{itemize}
These statements are in close analogy to the (abelian) spin-1 case, where it is the $m =0$ polarization that decouples or dominates in the limit $M_V\to 0$, depending on whether the corresponding vector current that $V$ couples to is conserved or not. 

From the above it is already clear that (boosted) spin-2 particles produced in DM annihilation (or decay) will be polarized to some degree, and hence lead to non-flat photon and neutrino spectra.
It is relatively easy to cook-up a model that dominantly produces $T$ with $m=0$ in DM annihilation processes, simply by coupling $T$ to a non-conserved source. The real challenge is therefore to find models that produce the other $T$ polarizations (necessarily involving a conserved source, henceforth the energy--momentum tensor of the DM particle). The $m=\pm 2$ modes are of particular interest because they will lead to highly peaked gamma-ray spectra (see left plot of Fig.~\ref{fig:BoxesSpin2} and note that $T\to \gamma\gamma$ only has non-zero coefficients $C_{\pm 2}$, as shown in Tab.~\ref{table:TensorC}).

 We will now present several examples to illustrate our statements. In all cases we will take the source $\mathcal{T}_{\mu\nu}^X$ to be the energy--momentum tensor of a new particle $X$, either scalar, fermion or vector. Most of the relevant Feynman rules can be found in Ref.~\cite{Han:1998sg}.

\subsection{\texorpdfstring{$VV \to V T$}{VV -> VT}}

In order to produce all five polarizations when only one tensor is emitted in $s$-wave DM annihilations, it seems that rather high spins are required.
As such, we start our tour of models with vector DM. The ideal model for our purposes was proposed in Ref.~\cite{Hambye:2008bq}, as it contains stable spin-1 DM. Here, the dark sector has an $SU(2)$ gauge symmetry, spontaneously broken by a complex scalar doublet $\phi$,
\begin{align}
\L_V = -\tfrac14 V_{\mu\nu}^a V^{\mu\nu,a} + |D_\mu \phi|^2 - V_\mathrm{pot}(\phi,H)\,,
\end{align}
with field strength tensor $V_{\mu\nu}^a\equiv \del_\mu V_\nu^a - \del_\nu V_\mu^a + g \epsilon^{abc} V_\mu^b V_\nu^c$.
In absence of fermions charged under the $SU(2)$, an accidental custodial $SO(3)$ symmetry survives after symmetry breaking, which keeps the three massive vectors $V_\mu^{1,2,3}$ degenerate and stable. In unitary gauge, $\phi = (0, v_S + S)^T/\sqrt{2}$, the relevant interaction terms are
\begin{align}
\begin{split}
\L_V &= -\tfrac14 V_{\mu\nu}^a V^{\mu\nu,a} + \tfrac12 (\del_\mu S)^2 - V_\mathrm{pot}(S,h)\\
 &\quad+ \tfrac12 M_V^2 V_\mu^a V^{\mu,a} \left(1+\frac{S}{v_S}\right)^2 .
\end{split}
\end{align}
The vacuum expectation value can be expressed in terms of the gauge coupling $g$ and vector mass as $v_S \equiv 2M_V/g$.
The Brout--Englert--Higgs-like scalar $S$ remaining in unitary gauge provides a portal to the SM sector, as it mixes with the SM scalar $h$ through the scalar potential $V_\mathrm{pot}$~\cite{Hambye:2008bq}. 
From the Lagrangian $\L_V$ we can calculate the corresponding energy--momentum tensors $\mathcal{T}_{\mu\nu}^V$ and $\mathcal{T}_{\mu\nu}^S$ of the particles $V$ and $S$, which defines the coupling to the spin-2 particle $T$ via Eq.~\eqref{eq:fierz-pauli}.
This gives rise to the processes $VV\to VT$ and $VV\to ST$, which have a promising high number of spin-couplings to produce $T_{\pm 2}$.

\begin{figure}[t]
\includegraphics[width=0.4\textwidth]{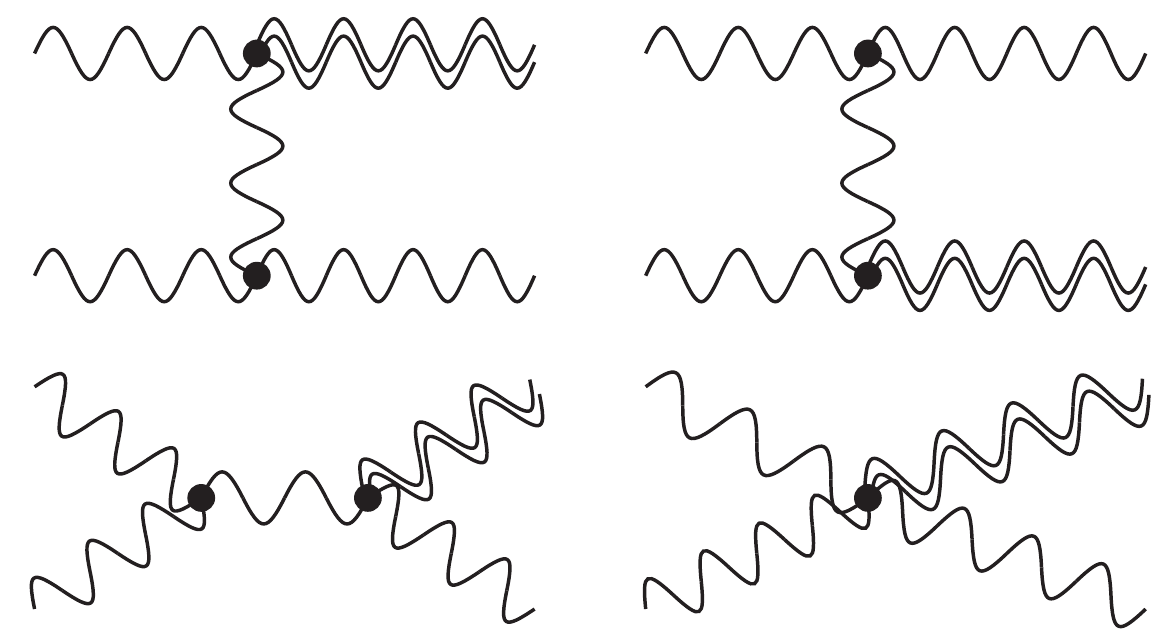}
\caption{
Feynman diagrams for $V V \to V T$, where the double-squiggly line stands for the spin-2 tensor $T$.
}
\label{fig:VVtoVT}
\end{figure}

We start our discussion with the process $VV\to VT$ (see Fig.~\ref{fig:VVtoVT}), which exists due to the non-abelian nature of the dark sector. The $s$-wave cross section for $V_a V_b\to V_c T$ is only non-vanishing for $a\neq b\neq c\neq a$ due to the custodial $SO(3)$ symmetry,
\begin{align}
\sigma v (V_a V_b\to V_c T) = \frac{g^2}{192\pi \Lambda^2}  u_{VVVT} (r_T^2) \,,
\label{eq:VVtoVT}
\end{align}
where $r_T\equiv M_T/M_V$ and
\begin{align}
\begin{split}
u_{VVVT} (x) &\equiv \frac{27 + 15 x -29 x^2 -15 x^3 + 2 x^4}{27(1-x/3)^2} \\
&\quad \times \left[1-10 x + 9 x^2\right]^{\frac12} .
\end{split}
\end{align}
The limit $M_T\to 0$ is unproblematic due to the conserved source tensor, in particular $u_{VVVT} (0) = 1$.\footnote{We cross checked our calculation by comparing the relativistic limit to $\sigma (gg \to g T)$ from Ref.~\cite{Mirabelli:1998rt}.} 
The helicity branching ratios 
\begin{align}
\Br_{m} \equiv \frac{\sigma v (VV\to V T_m)}{\sum_{m'}\sigma v (VV\to V T_{m'})} \,,
\end{align}
are quite simple,
\begin{align}
\Br_{0} : \Br_{\pm 1} : \Br_{\pm 2} = 6 : 5 r_T^2 : 2 r_T^4\,.
\end{align}
For small $M_T$, we find that the $m =\pm 1$ helicities decouple with $r_T^2$, as expected for a conserved source. The $m =\pm2$ modes decouple even faster, making $m =0$ the dominant polarization in the entire the parameter space (since $r_T< 1$ due to kinematics). 

\begin{figure}[t]
\includegraphics[width=0.45\textwidth]{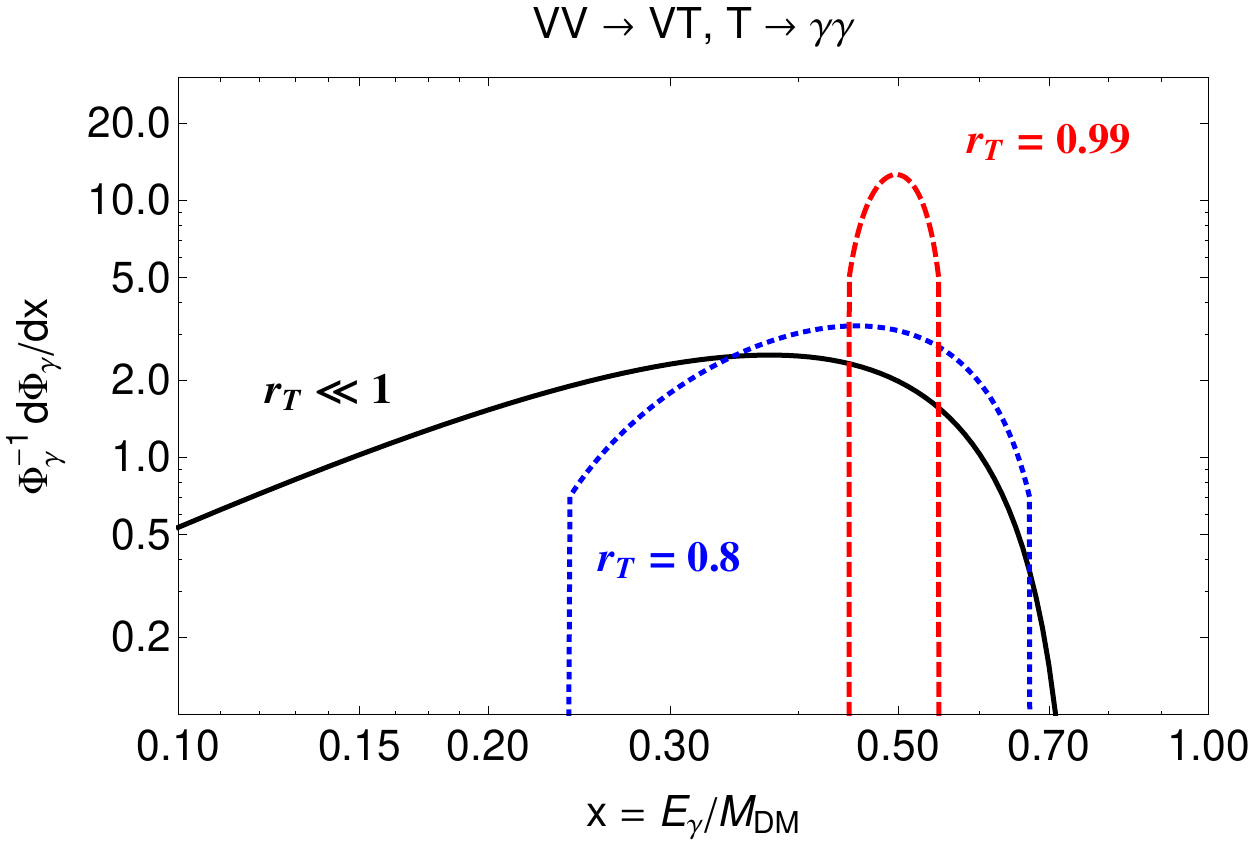}\\
\includegraphics[width=0.45\textwidth]{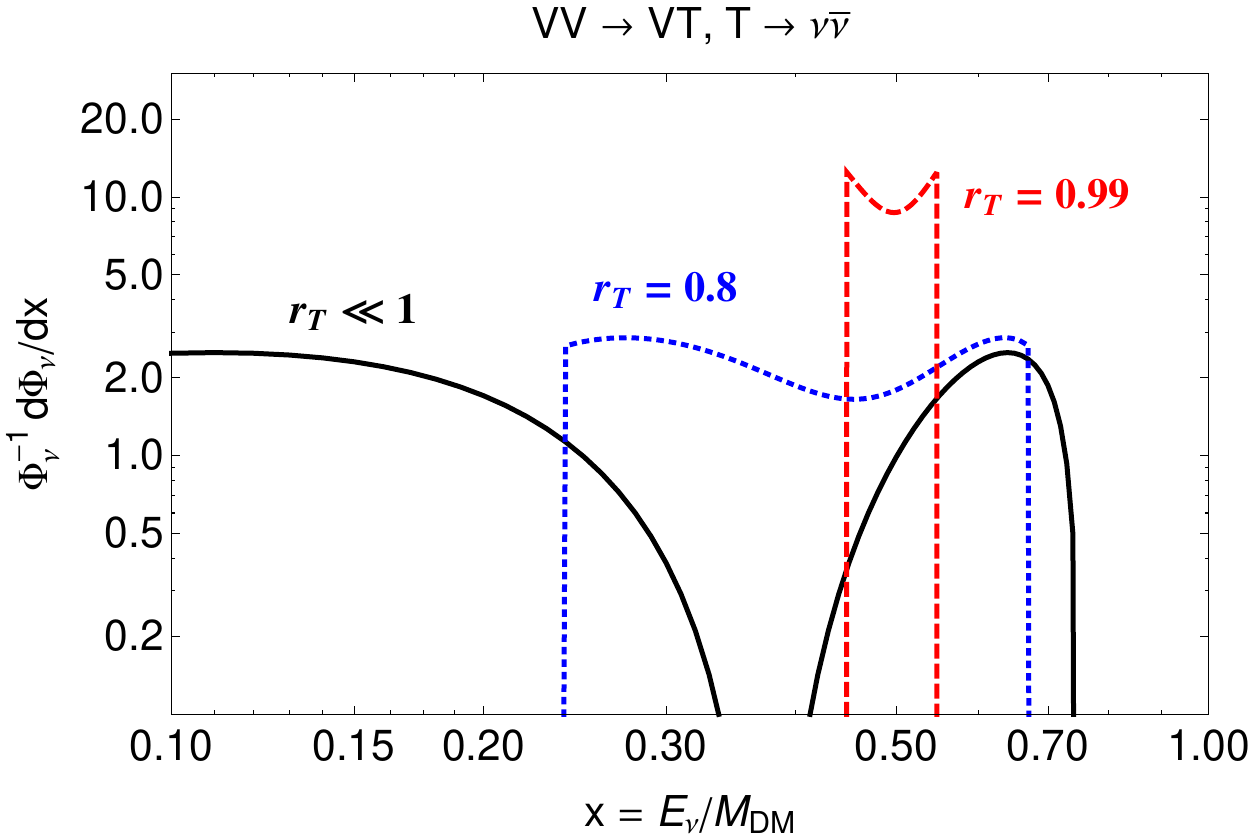}
\caption{
Differential flux from $VV\to VT$, followed by $T\to \gamma\gamma$ (upper panel) or $T\to \nu\bar\nu$ (lower panel).
}
\label{fig:VVtoVTspectrum}
\end{figure}

Together with the branching ratios of $T_m$ into photons and neutrinos (see Tab.~\ref{table:TensorC}), we can calculate the flux and spectral shape of our processes analogous to Eq.~\eqref{eq:neutrinoflux}. In Fig.~\ref{fig:VVtoVTspectrum} we give the photon and neutrino spectrum (identical to anti-neutrino spectrum) for various values of $r_T$. 
The width of the ``box'' is the same for both final states, but the spectral shape within the box is completely different for neutrinos and photons as a result of their different spins.
A more detailed analysis of these spectra will have to wait until Sec.~\ref{sec:spectra}, for now we merely point out the interesting double-peak shape of the neutrino flux.

The features from Fig.~\ref{fig:VVtoVTspectrum} will unavoidably be accompanied by actual lines from the $s$-channel processes $VV\to T\to \gamma\gamma,\nu\bar\nu$. The cross sections for monochromatic photons~\cite{Han:2015cty} and neutrinos are
\begin{align}
\sigma v (V_a V_a \to \gamma\gamma) &= \frac{8}{9\pi} \frac{M_V^4}{\Lambda^4}\frac{M_V^2}{(4M_V^2-M_T^2)^2 + \Gamma_T^2 M_T^2}\,,\\
\sigma v (V_a V_a \to \nu\bar \nu) &= \frac{2}{9\pi} \frac{M_V^4}{\Lambda^4}\frac{M_V^2}{(4M_V^2-M_T^2)^2 + \Gamma_T^2 M_T^2}\,,
\end{align}
which are parametrically suppressed by $M_V^2/\Lambda^2$ compared to the single tensor emission of Eq.~\eqref{eq:VVtoVT}.\footnote{Here we consider neutrinos to be Weyl or Majorana spinors, reducing the cross section (and $T\to\nu\bar\nu$ branching ratio) by a factor $1/2$ compared to a Dirac neutrino~\cite{Bonora:2014qla}.} Note that due to the kinematics, these processes give lines at $E_{\gamma,\nu} = M_V$, i.e.~at \emph{higher} energies than the spectral features of Fig.~\ref{fig:VVtoVTspectrum}. 

\subsection{\texorpdfstring{$VV \to S T$}{VV -> ST}}

Let us now discuss $V_a V_b\to ST$ within the same model~\cite{Hambye:2008bq} from above, which is only non-vanishing for $a=b$ due to the $SO(3)$ symmetry. The $s$-wave cross section takes the form
\begin{align}
\sigma v (V_a V_a\to S T) =\frac{11}{1728\pi} \frac{g^2}{\Lambda^2} u_{VVST} (r_T^2,r_S^2) \,,
\label{eq:VVtoST}
\end{align}
with $r_{T,S} \equiv M_{T,S}/M_V$ and $u_{VVST} (x,y)$ defined in Eq.~\eqref{eq:uVVST} in the appendix. The limit $M_T\to 0$ is again unproblematic due to the conserved source tensor and we have $u_{VVST} (0,0) = 1$. The cross section is generically of the same order as $VV\to VT$ if the scalar is kinematically accessible. 
In case the mixing angle $\alpha$ between $S$ and the SM scalar $h$ is not zero, one has to multiply the above formula by $\cos^2\alpha$ or $\sin^2\alpha$ and replace $M_S = r_S M_V$ by the appropriate mass.
The helicity branching ratios are given in full form in Eqs.~\eqref{eq:BrVVST1}--\eqref{eq:BrVVST2} in the appendix, which reduce to
\begin{align}
\Br_0 &\simeq 1-\frac{48 r_T^2}{44-12 r_S^2+r_S^4}\,,\\
\Br_1 &\simeq \frac{24 r_T^2}{44-12 r_S^2+r_S^4}\,,\\
\Br_2 &\simeq\frac{384 r_T^4}{\left(4-r_S^2\right)^2 \left(44-12 r_S^2+r_S^4\right)}\,,
\end{align}
for $r_T\ll 1$.
Qualitatively, these give the same $M_T\to 0$ decoupling pattern that we found above for $VV\to VT$. 
Even for large $r_T$ it is always the $m=0$ helicity mode that dominates, so the flux will be similar to that of $VV\to VT$ in Fig.~\ref{fig:VVtoVTspectrum}.

\subsection{\texorpdfstring{$F\bar F\to V T$}{FF -> VT}}
\label{sec:FFtoVT}

With the above vector DM model we found examples of how to dominantly produce the $m=0$ tensor polarization. Let us try to find examples where $m=0$ is subleading.
Fermion DM annihilating into one vector boson and one tensor $T$ should allow for the production of all five $T$ helicities in the $s$ wave, so is a worthwhile example to study here.
A simple model consists of one massive Dirac fermion~$F$ charged under a $U(1)'$ with massive gauge boson $V$,
\begin{align}
\begin{split}
\L_F &= -\tfrac14 V_{\mu\nu} V^{\mu\nu} + \tfrac{M_V^2}{2} V_\mu V^{\mu} +\overline{F} \left(i \slashed{\del} - g \slashed{V} -M_F \right) F\,.
\end{split}
\end{align}
For our purposes it suffices to give a St\"uckelberg mass to $V$, seeing as we are not interested in the effects of spontaneously breaking the $U(1)'$.
The stability of $F$ can in either case be ensured by an unbroken discrete subgroup of $U(1)'$, for example by giving $F$ an even $U(1)' = U(1)_{B-L}$ charge~\cite{Batell:2010bp,Heeck:2015pia}. 
Coupling $T_{\mu\nu}$ to the energy--momentum tensor $\mathcal{T}_{\mu\nu}^F$ derived from $\L_F$~\cite{Han:1998sg} allows us to calculate the unpolarized cross section for the process $F\bar F\to V T$ in the non-relativistic ($s$-wave) limit,
\begin{align}
\sigma v (F\bar F\to V T) &=\frac{1}{48\pi} \frac{g^2}{\Lambda^2}\, u_{FFVT} (r_T^2,r_V^2) \,,
\label{eq:FFtoVT}
\end{align}
where $r_{T,V} \equiv M_{T,V}/M_F$ and the expression for $u_{FFVT}(x,y)$ is given in Eq.~\eqref{eq:uFFVT}.\footnote{To cross check our calculation we also considered the relativistic limit and compared to $\sigma (e^+e^- \to \gamma T)$ from Ref.~\cite{Mirabelli:1998rt}.}
We stress that both limits $M_T\to0$ and $M_V\to0$ lead to finite rates due to the coupling to conserved sources and yield $u_{FFVT} (0,0) = 1$.

\begin{figure}[t]
\includegraphics[width=0.45\textwidth]{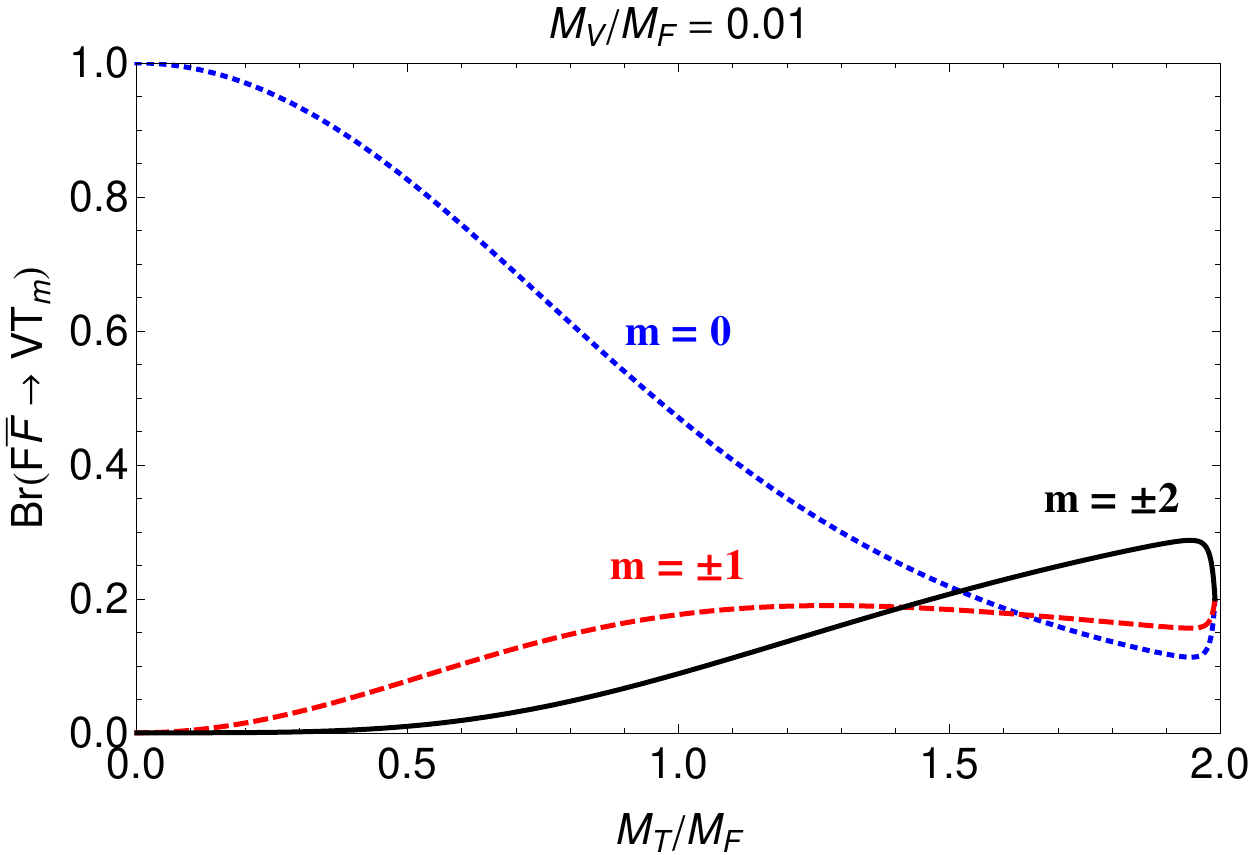}
\caption{
Branching ratios of the different spin-2 helicity states in $F\bar F\to V T_m$ for $M_V/M_F = 0.01$.
}
\label{fig:FFtoVTspectra}
\end{figure}

\begin{figure}[t]
\includegraphics[width=0.45\textwidth]{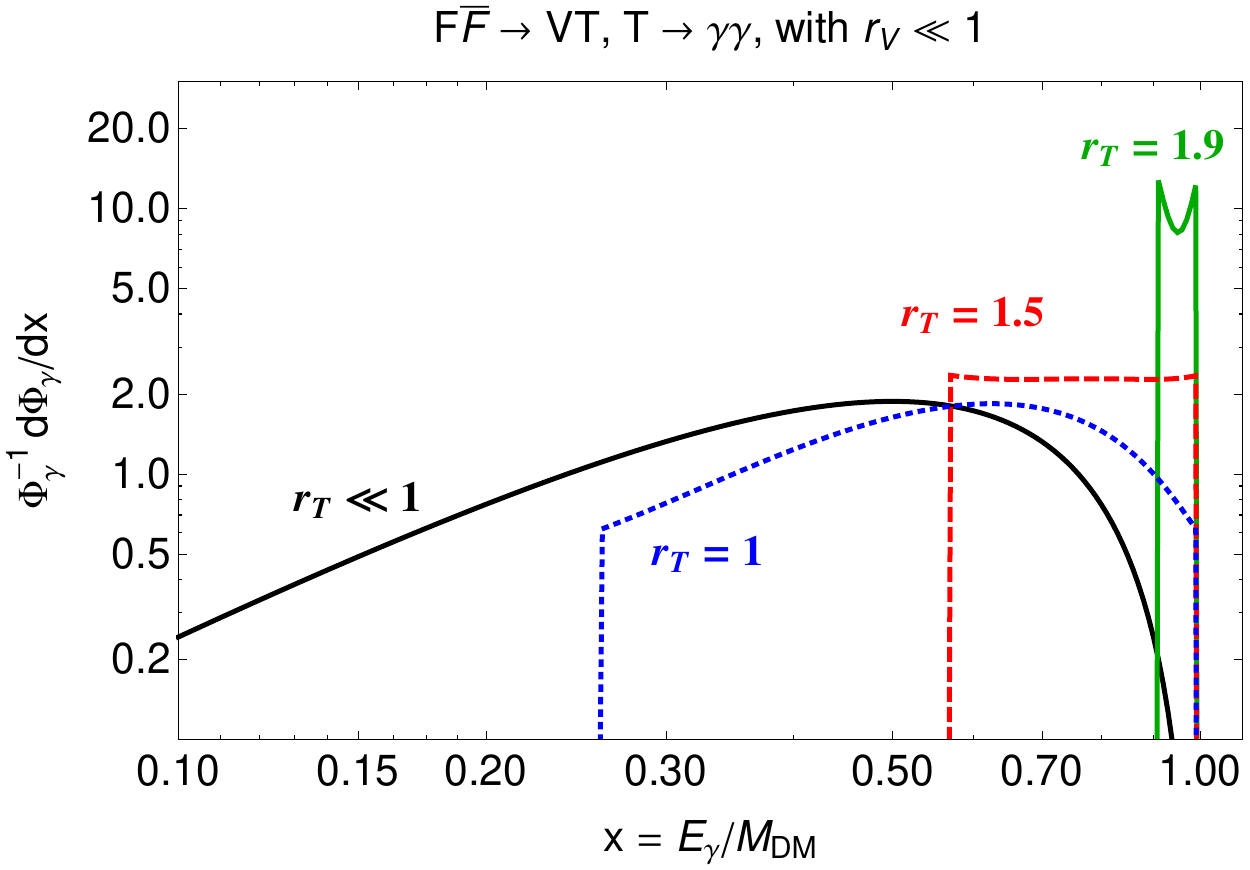}\\
\includegraphics[width=0.45\textwidth]{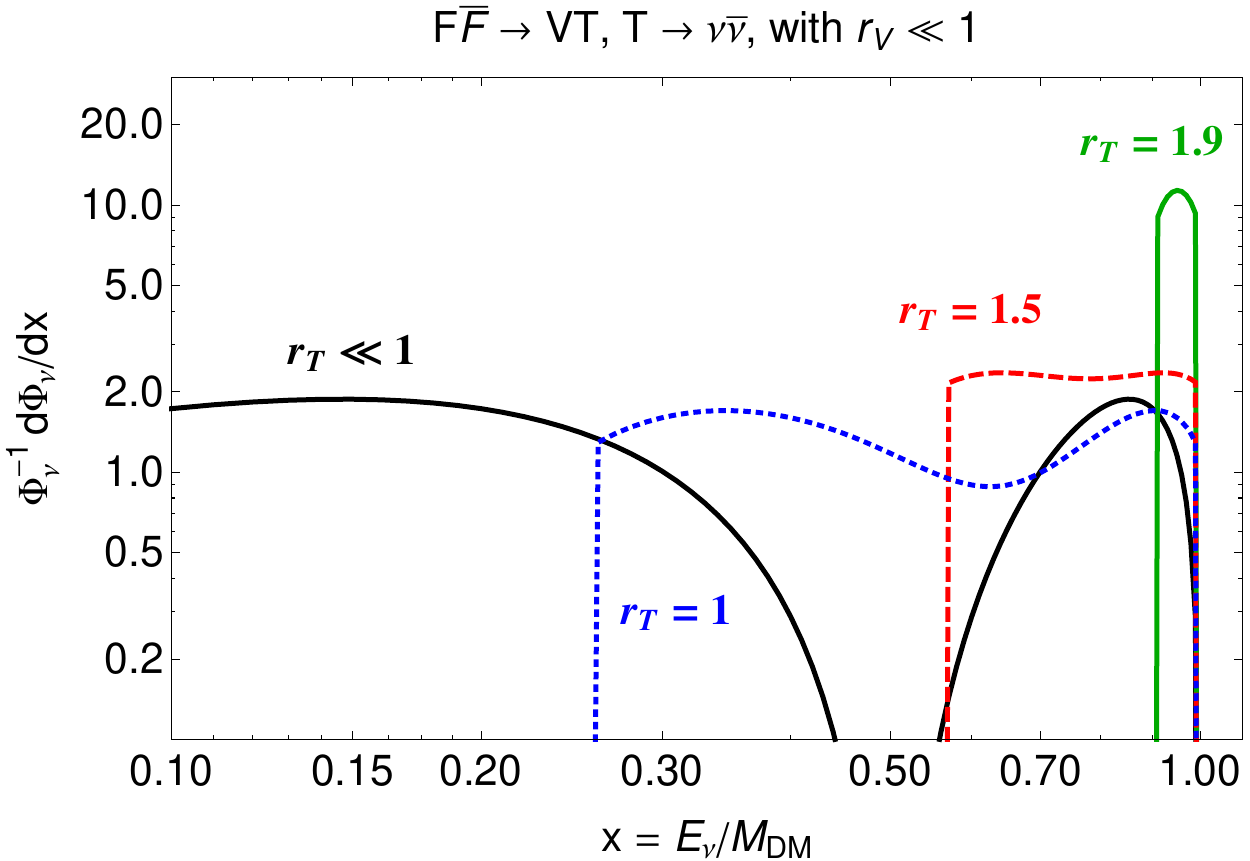}
\caption{
Differential flux from $F\bar F\to VT$, followed by $T\to \gamma\gamma$ (upper panel) or $T\to\nu\bar\nu$ (lower panel).
}
\label{fig:FFtoVTphotonspectrum}
\end{figure}

The full expressions for $\Br_{m} $ are given in Eqs.~\eqref{eq:BrFFVT1}--\eqref{eq:BrFFVT2}, let us simply focus on the limit $r_T\ll 1$:
\begin{align}
\Br_{0} &\simeq 1-\frac{6 r_T^2}{8+r_V^2}\,,\\
\Br_{\pm 1} &\simeq\frac{3 r_T^2}{8+r_V^2}\,,\\
\Br_{\pm 2} &\simeq \frac{24 r_T^4}{\left(4-r_V^2\right)^2 \left(8+r_V^2\right)}\,.
\end{align}
This is the same qualitative behavior we found for vector DM in the previous sections.
For highly boosted $T$, we thus find again dominantly $m =0$. Let us consider the opposite case of rather slow $T$. With Eqs.~\eqref{eq:BrFFVT1}--\eqref{eq:BrFFVT2} one can show that $\Br_m \to 1/5$ for all $m$ in the limit $M_T \to 2M_F -M_V$ with $M_V\neq 0$, corresponding to an unpolarized $T$ at rest. 
An interesting effect occurs however for $r_V \ll 1$,\footnote{We checked that this limit gives the same result as a direct calculation with a massless $V$, i.e.~the longitudinal $V$ mode completely decouples from the process.} which yields
\begin{align}
\Br_{0} : \Br_{\pm 1} : \Br_{\pm 2} \simeq 16 : 6 r_T^2 : 3 r_T^4\,.
\end{align}
Since kinematics only restrict $r_T + r_V < 2$, the $m =0$ mode is no longer dominant for $r_T \gtrsim 1.5$, as $m =\pm2$ gets enhanced (see Fig.~\ref{fig:FFtoVTspectra}). This effect is amplified if we properly set $M_V = 0$, as this allows $T$ to be polarized even if produced (almost) at rest. Combining the above with the $T$ decay spectra from Tab.~\ref{table:TensorC}, we can obtain the resulting gamma-ray and neutrino spectra from $F\bar F\to V (T\to\gamma\gamma,\nu\bar\nu)$ (Fig.~\ref{fig:FFtoVTphotonspectrum}). While for most of the parameter space we obtain the $m =0$ spectrum, this is not necessarily  the case for $M_V\ll M_F \lesssim M_T$, as argued above. However, in this part of the parameter space the box becomes increasingly narrow, making it hard to determine the spectral shape within for realistic detector resolutions. We will come back to this issue in Sec.~\ref{sec:spectra}.

For completeness, we give the processes that yield real monochromatic photons~\cite{Han:2015cty} or neutrinos
\begin{align}
\begin{split}
\sigma v (F\bar F \to \gamma\gamma) \simeq \frac{v^2}{12\pi} \frac{M_F^4}{\Lambda^4}\frac{M_F^2}{(4M_F^2-M_T^2)^2 + \Gamma_T^2 M_T^2}\,,\\
\sigma v (F\bar F \to \nu\bar \nu) \simeq \frac{v^2}{48\pi} \frac{M_F^4}{\Lambda^4}\frac{M_F^2}{(4M_F^2-M_T^2)^2 + \Gamma_T^2 M_T^2}\,.
\end{split}
\label{eq:FFtonunu}
\end{align}
Both are $p$-wave suppressed and, of course, of higher order in the coupling constant $1/\Lambda$ compared to the single tensor emission $F\bar F\to V T$ of Eq.~\eqref{eq:FFtoVT}.

\subsection{\texorpdfstring{$F F \to T T$}{FF -> TT}}

\begin{figure}[t]
\includegraphics[width=0.4\textwidth]{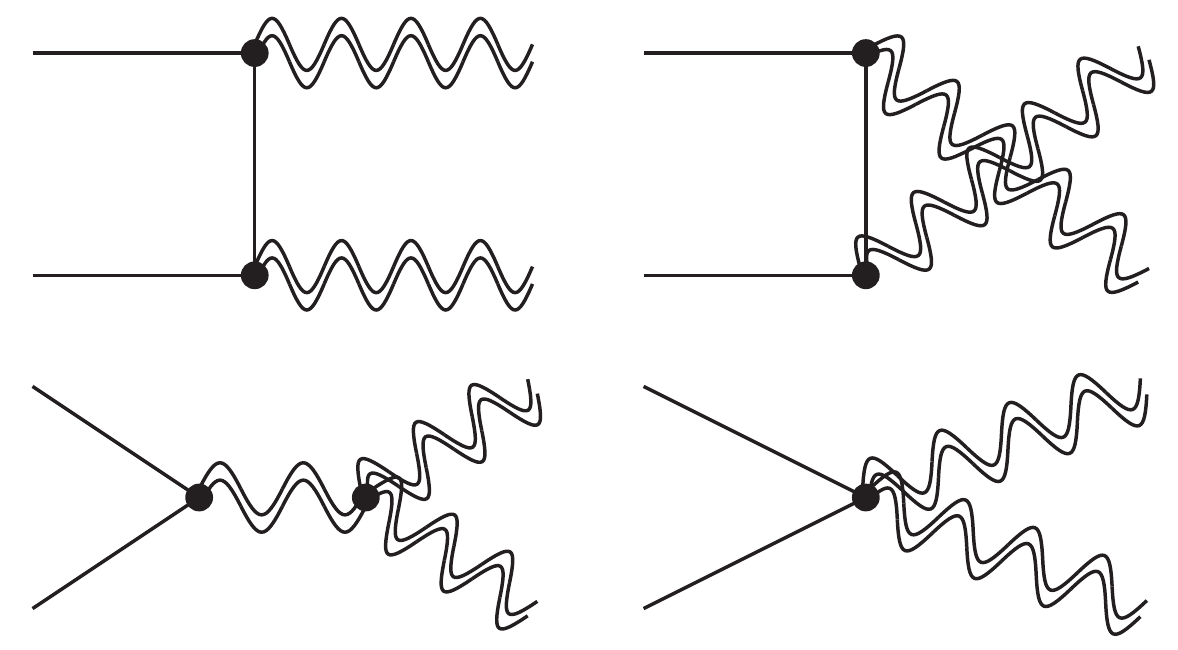}
\caption{
Feynman diagrams for $F F \to T T$, where the double-squiggly line stands for the spin-2 tensor $T$ and the fermion is Majorana.
}
\label{fig:FFtoTT}
\end{figure}

All processes considered so far dominantly gave  rise to the $m=0$ tensor polarization. Let us give an example for a process where the $m=0$ mode is \emph{forbidden}, making the $m=\pm 2$ polarizations relevant. For this, we consider a \emph{Majorana} fermion $F$ and calculate the rate $F F\to TT$ in the $s$ wave. In this case, the transition amplitude  can be cast as  ${\cal M} = \bar{v} {\cal F} u$, where $v$ and $u$ are the spinors associated to the initial fermion describing arbitrary spin states. We are however only interested in the state of total spin $S=0$, since the ones associated to $S=1$ do not exist for a pair of identical fermions in the $s$-wave configuration. In order to isolate the amplitude for the $S=0$ state we follow Refs.~\cite{Kuhn:1979bb,Bergstrom:1997fh} and calculate it by means of 
\begin{align}
\begin{split}
{\cal M}^{S=0} =- \frac{1}{\sqrt2}
&\text{Tr} \left\{ {\cal F} \left(\slashed{k}+M_\text{DM}\right) \gamma^5 \right\} ,
\end{split}
\label{eq:P0}
\end{align}
where $k= (M_\text{DM},0,0,0)$ is the DM momentum in the non-relativistic limit. Incidentally, this procedure shows that the $s$-channel diagram (see~Fig.~\ref{fig:FFtoTT}) does not contribute to the annihilation amplitude because the vertex $FFT$, when inserted in Eq.~\eqref{eq:P0}, gives zero. Consequently, for this specific non-relativistic $s$-wave cross section it is not necessary to include the cubic tensor self-interactions, which is a non-trivial ingredient of theories with massive spin-2 particles.  The nevertheless required contact-interaction vertex $FFTT$ can be found, e.g.~in Ref.~\cite{Holstein:2006bh}.
This procedure yields the $s$-wave cross section
\begin{align}
\sigma v (F F \to TT) = \frac{m_F^2}{64\pi\Lambda^4} (1-r_T^2)^{\frac32} \left(1+\frac{r_T^4/4}{(2-r_T^2)^2}\right) .
\end{align}
While this is of the same $1/\Lambda$ order as the cross sections into monochromatic neutrinos and photons, the latter are $p$ wave and hence additionally suppressed.

The $s$-wave Majorana fermions couple to an initial $J^{PC}=0^{-+}$ state, so the two tensor polarizations are always the same, with simple ratios
\begin{align}
\Br_{0,0} : \Br_{\pm 1,\pm 1} : \Br_{\pm 2,\pm 2} = 0 :  r_T^4 :  (4- 2 r_T^2)^2\,.
\end{align}
The $m=\pm 1$ modes again decouple for $r_T\to 0$; the qualitatively new feature in this process is the vanishing $m=0$ amplitude, even for massive spin-2 particles. This is in fact a selection rule ensured by symmetry: the initial state has negative parity, so the final state must have \emph{odd} orbital angular momentum $L$. For Majorana particles in the $s$-wave $J = L+S = 0$, which  requires the two final-state tensors to couple to an \emph{odd} spin. Since odd spin states in boson--boson couplings are antisymmetric, we obtain $\Br_{0,0} =0$. Notice that this argument does not apply to the states with $m=\pm2$ because when they are produced, the two tensors in the final state have opposite spin components.

\begin{figure}[t]
\includegraphics[width=0.45\textwidth]{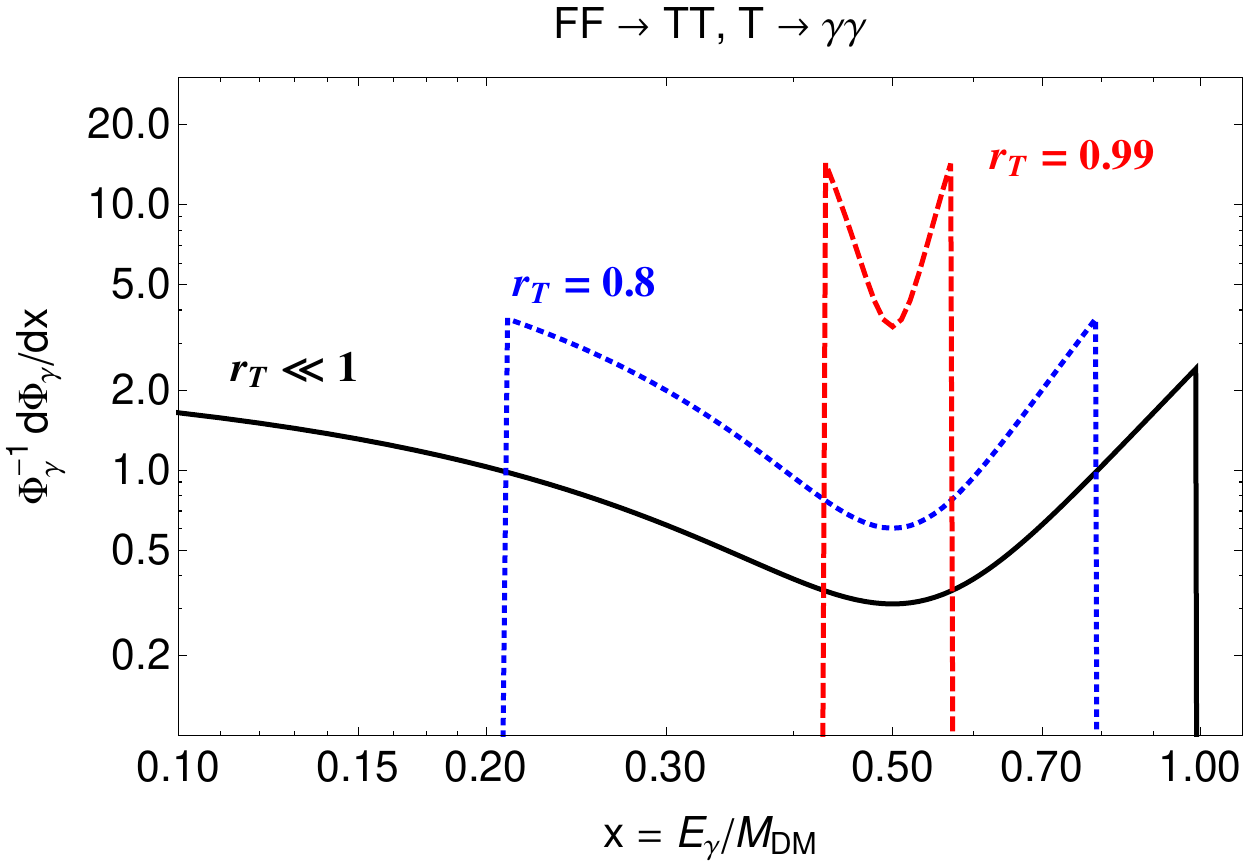}\\
\includegraphics[width=0.45\textwidth]{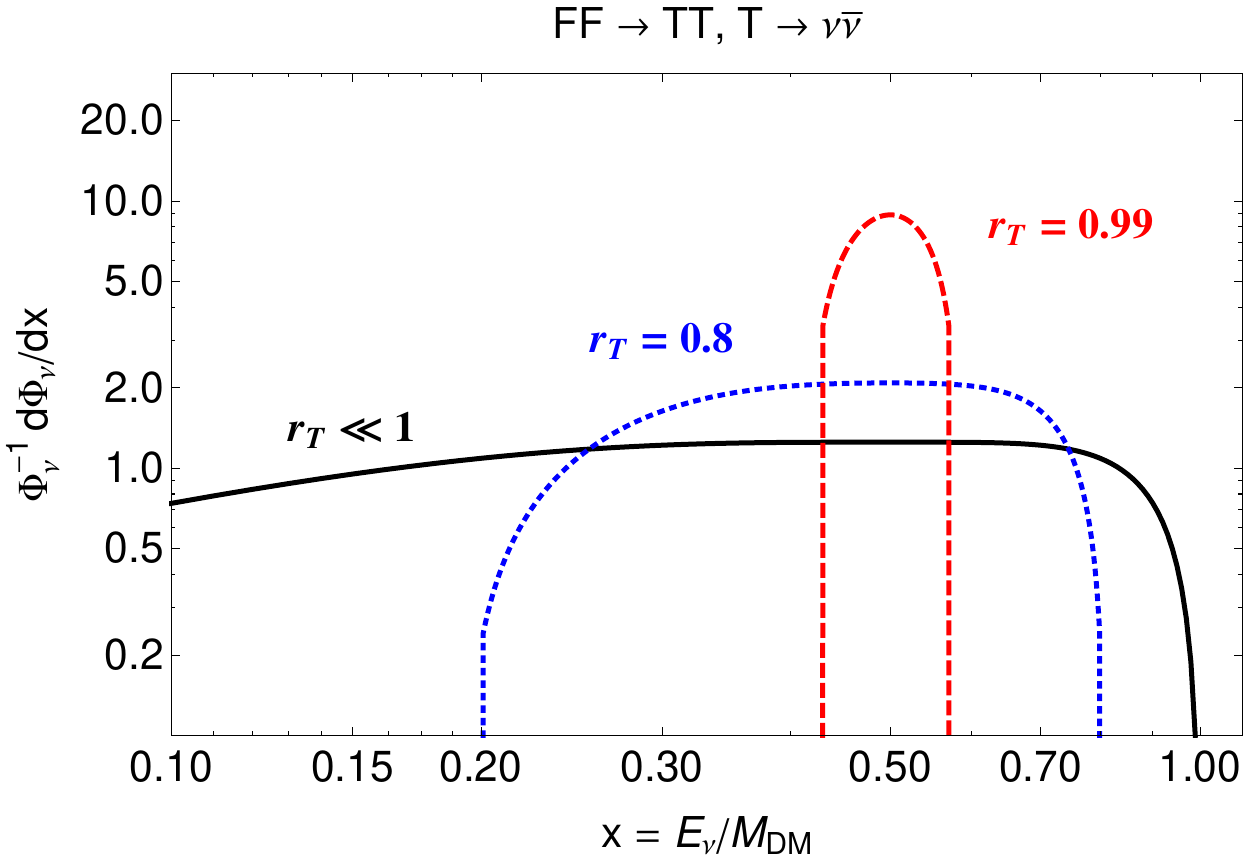}
\caption{
Differential flux from $F F\to TT$, followed by $T\to \gamma\gamma$ (upper panel) or $T\to \nu\bar\nu$ (lower panel).
}
\label{fig:FFtoTTspectrum}
\end{figure}

The resulting spectral shapes for gamma rays and neutrinos are shown in Fig.~\ref{fig:FFtoTTspectrum}. Due to the dominant $m=\pm 2$ polarization of the tensor, it is the photon spectrum that shows a double peak, compared to the spectra of $VV\to VT$ in Fig.~\ref{fig:VVtoVTspectrum}.

\section{Spectra from spin-2 mediators}
\label{sec:spectra}

\begin{figure*}[t]
\includegraphics[trim=0cm 0cm 0cm 0cm,clip,height=0.33\textwidth]{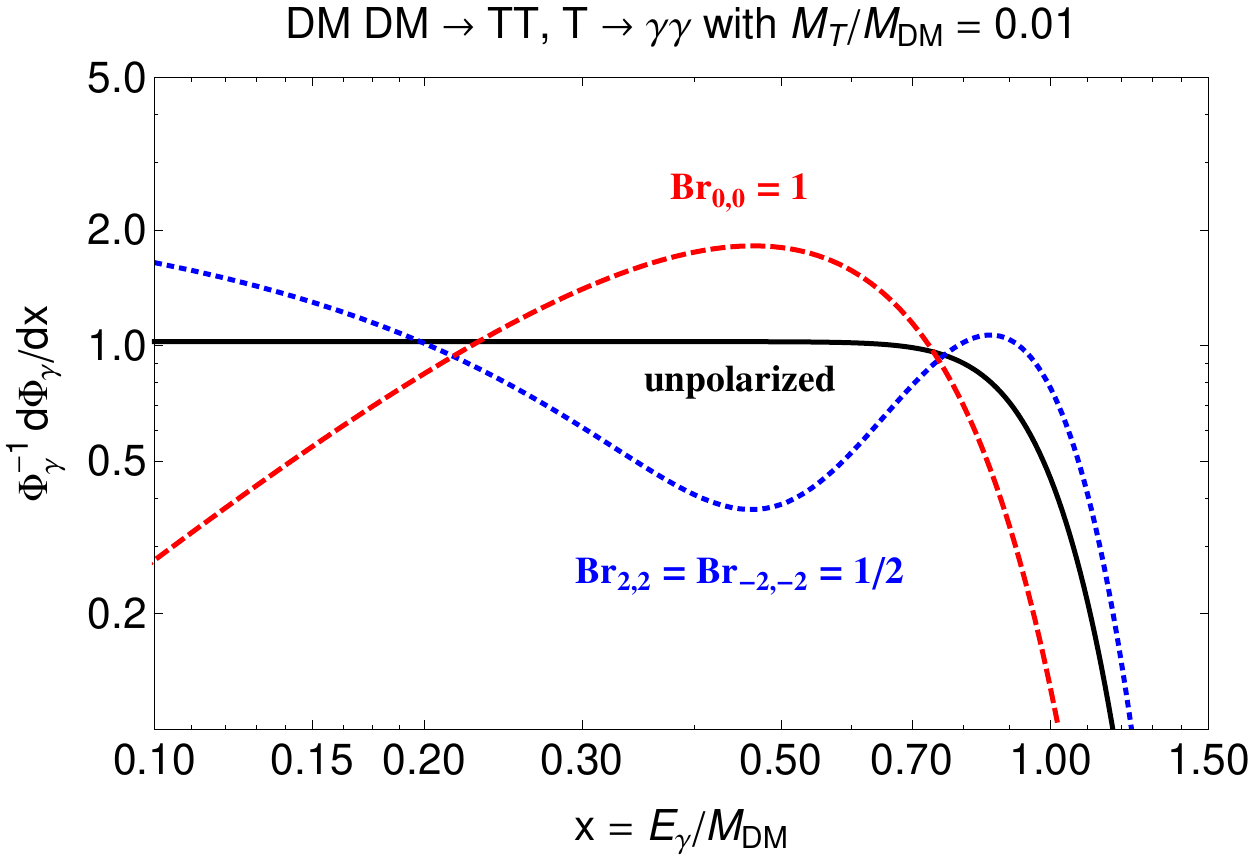}
\includegraphics[trim=0.7cm 0cm 0cm 0cm,clip,height=0.33\textwidth]{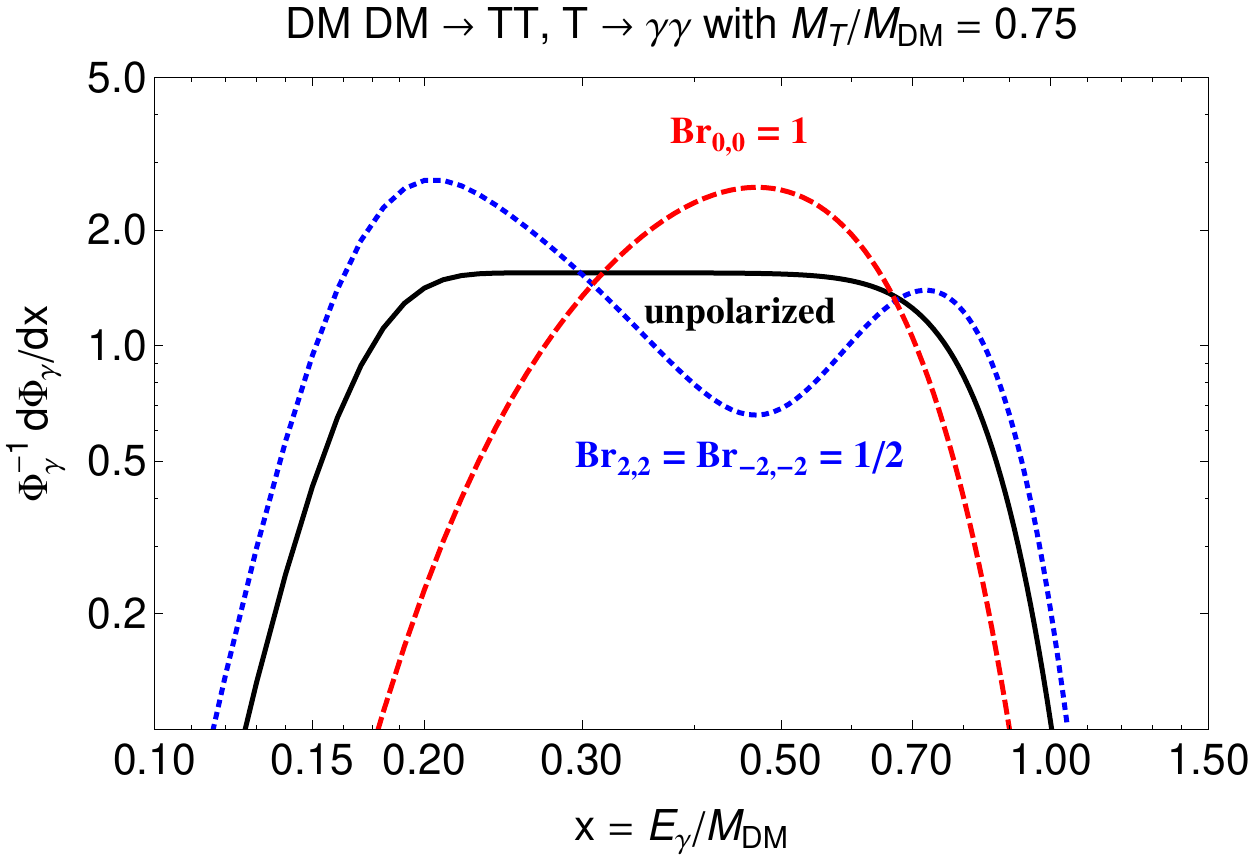}
\includegraphics[trim=0cm 0cm 0cm 0cm,clip,height=0.33\textwidth]{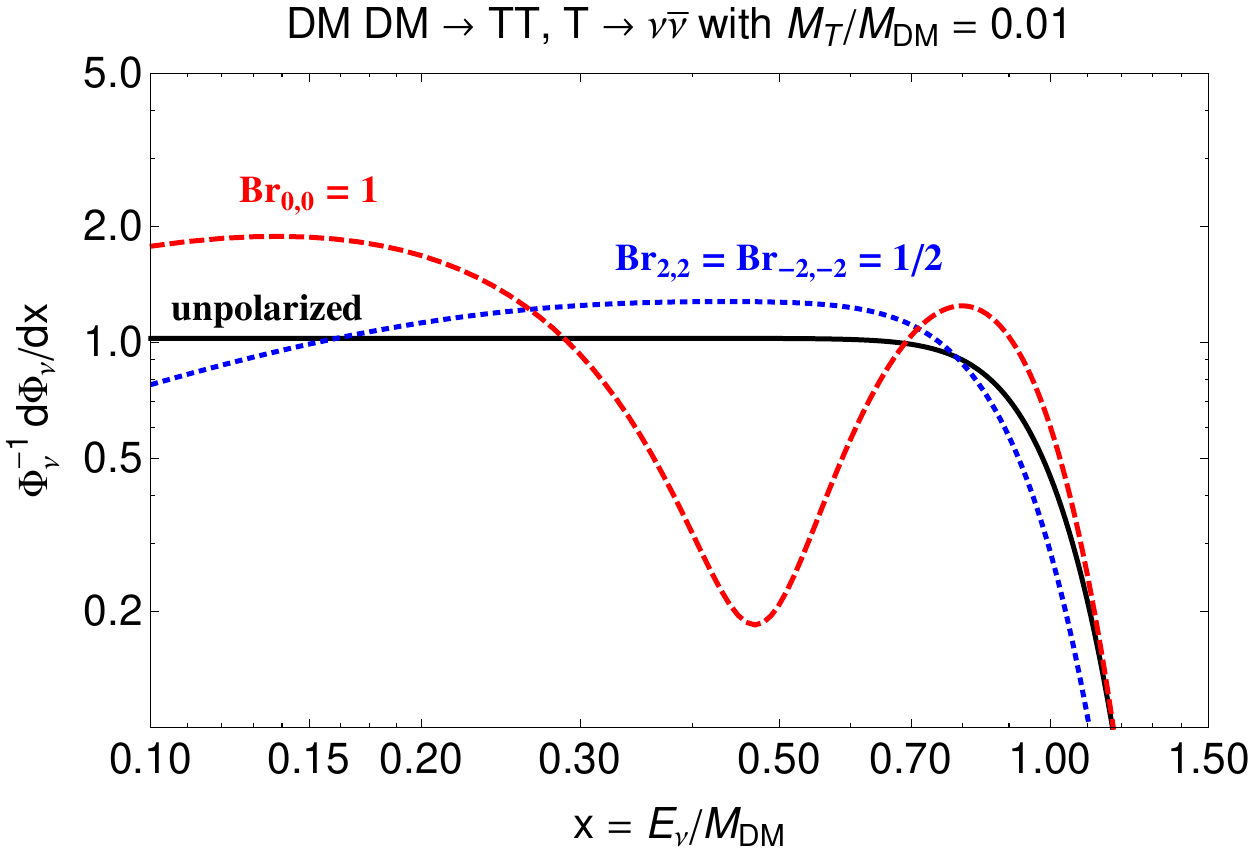}
\includegraphics[trim=0.7cm 0cm 0cm 0cm,clip,height=0.33\textwidth]
{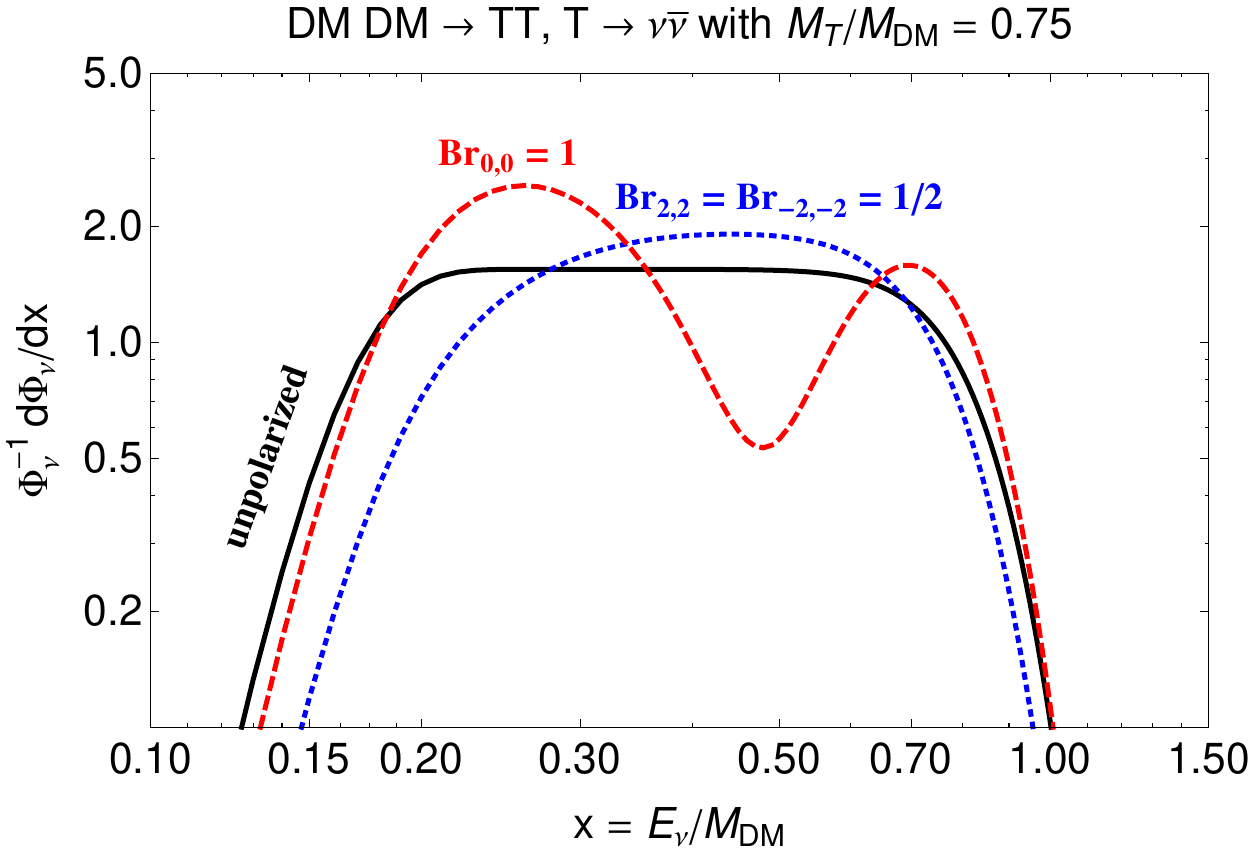}
\caption{
Differential flux expected from DM annihilation into a pair of spin-2 particles $T$ subsequently decaying into $\gamma\gamma$ (upper panels) or $\nu\bar\nu$ (lower panels), assuming an energy resolution of $15\%$ and different $T$ polarizations. The unpolarized spectrum is the same as the flat box of a spin-0 particle. The tensor mass is given by $M_T/M_\text{DM} = 0.01$ and $0.75$ in the left and right panels, respectively.
}
\label{fig:DMDMtoTT}
\end{figure*}

As long as we specify branching ratios into the different polarizations, the full generality of Eq.~\eqref{eq:Aflux} allows us to calculate model-independent spectra when the DM annihilates into  spin-2 particles coupled to the energy--momentum tensor. In fact, we have learned in the previous section that we typically have two cases: either the $m=0$ polarization dominates over the others or it is forbidden by a selection rule and the $m=\pm2$ states prevail.

Taking this as a motivation, we consider the process $\text{DM DM} \to TT$ assuming $L=J=0$ for the initial state and that the branching ratios are such that either $\Br_{0,0}= 1$ or $\Br_{2,2} = \Br_{-2,-2} = 1/2$. Then we let the tensor particles decay into a pair of photons or neutrinos and calculate the corresponding differential spectra using Eq.~\eqref{eq:Aflux}, taking $r_T= 0.01  $ and $r_T = 0.75$ as benchmark values. The resulting plots are shown in the left and right panels of Fig.~\eqref{fig:DMDMtoTT}, respectively. In order to account for the finite detector resolution, the spectra are convoluted with a Gaussian distribution of 15\% of the energy, a value which is at reach of current gamma-ray and neutrino telescopes~\cite{Abramowski:2013ax,Aartsen:2013vja}.
In the plots we have also included a scenario in which the spin-2 particles are produced unpolarized, which mimics the case in which the particles produced in the annihilation are scalars, as discussed in Sec.~\ref{sec:decay}. 

In the case of  $\Br_{0,0}= 1$, the neutrino spectra exhibit dips  and significantly deviate from the unpolarized flat box. In contrast, when the $m= \pm2$ states dominate, the spectra  are similar to the unpolarized ones, and consequently the resulting features are just the edges of the box.
This situation is somewhat reversed for gamma rays. There, the case $\Br_{0,0}= 1$ leads to spectra with no dips while the case when the polarizations $m=\pm2$ dominate gives rise to two spectral features clearly distinguishable from a flat box, with a dip at half of the spectrum.  Because the resolution increases with the energy, the height of those spectral features -- originally equal according to Fig.~\eqref{fig:BoxesSpin2} -- decreases for larger energies. 

\begin{figure*}[t]
\includegraphics[width=0.450\textwidth]{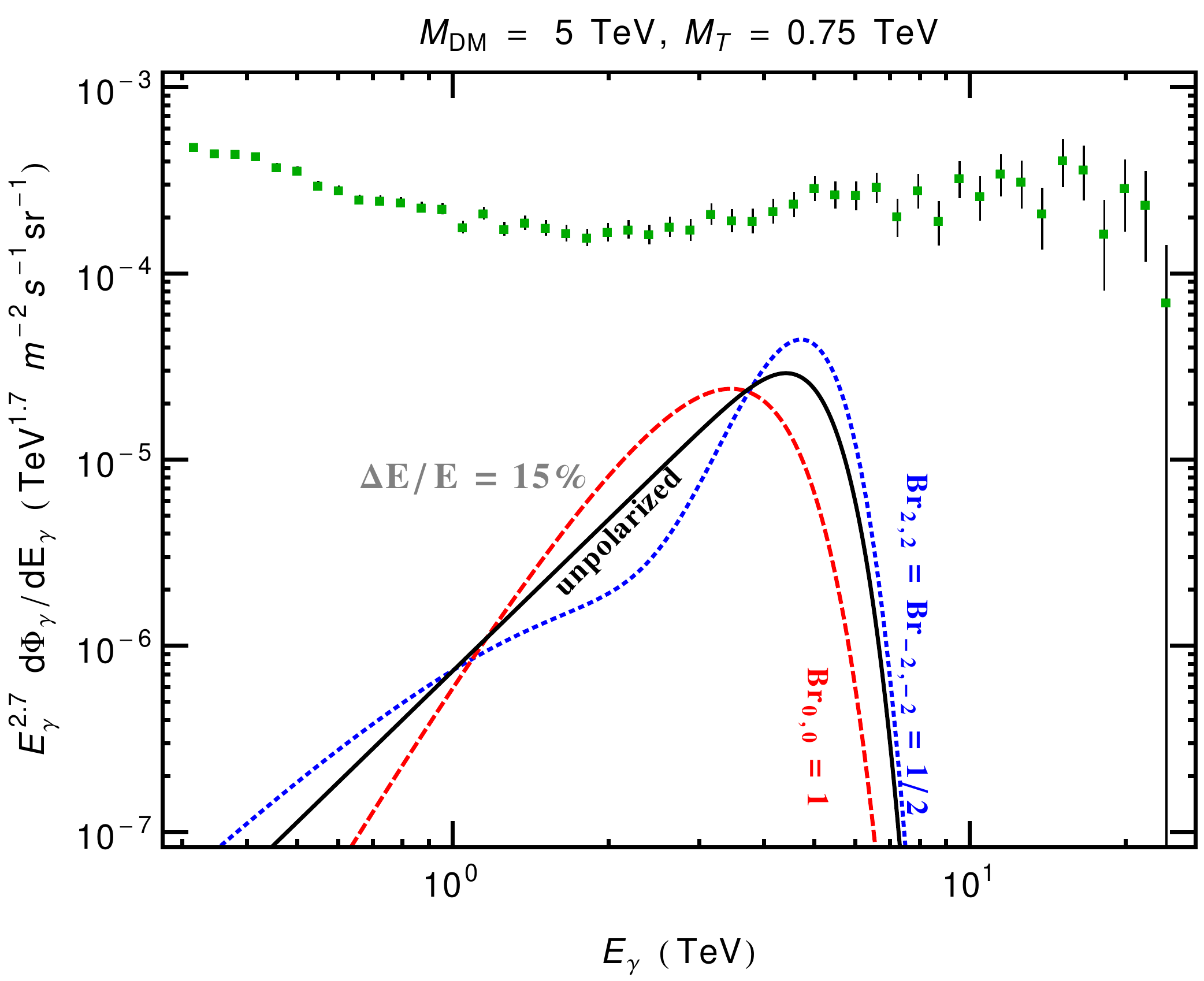}
\includegraphics[width=0.450\textwidth]{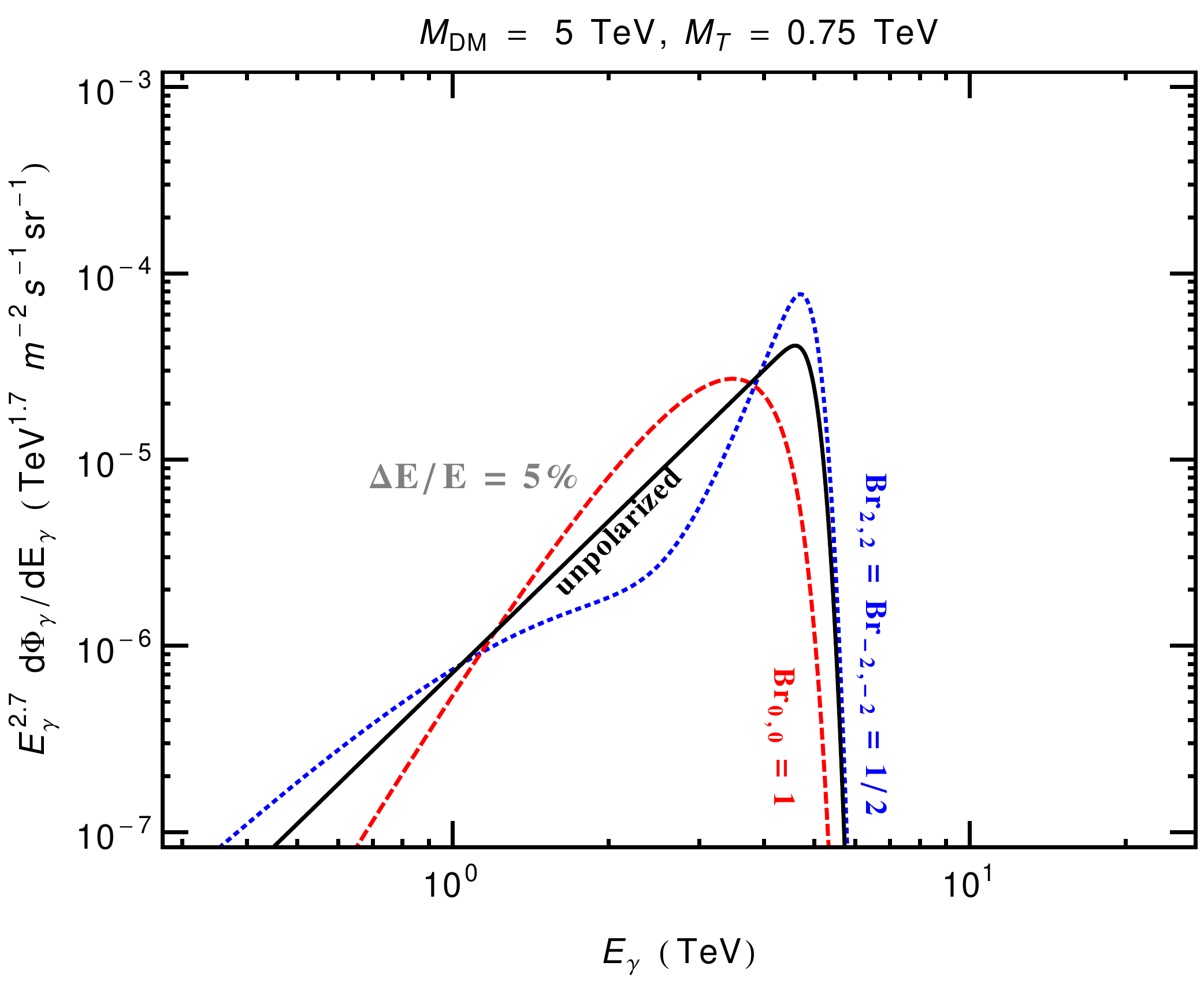}
\caption{ Left: Gamma-ray signal expected from DM  annihilation into a pair of spin-2 particles $T$ subsequently decaying into $\gamma\gamma$, along with the corresponding H.E.S.S.~data from galactic halo region~\cite{Abramowski:2013ax}. For illustrative purposes we choose $\Phi_\gamma =   \unit[3.5\times 10^{-6}]{m^{-2}s^{-1}sr^{-1}} $ for the overall DM signal. The unpolarized spectrum is the same as the flat box of a spin-0 particle. Right: Same DM spectrum as in the left panel but for an energy resolution of 5\% instead of 15\%, which could be achieved by future gamma-ray telescopes such as C.T.A.~\cite{Bernlohr:2012we}.
}
\label{fig:HESS}
\end{figure*}

\begin{figure*}[t]
\includegraphics[width=0.45\textwidth]{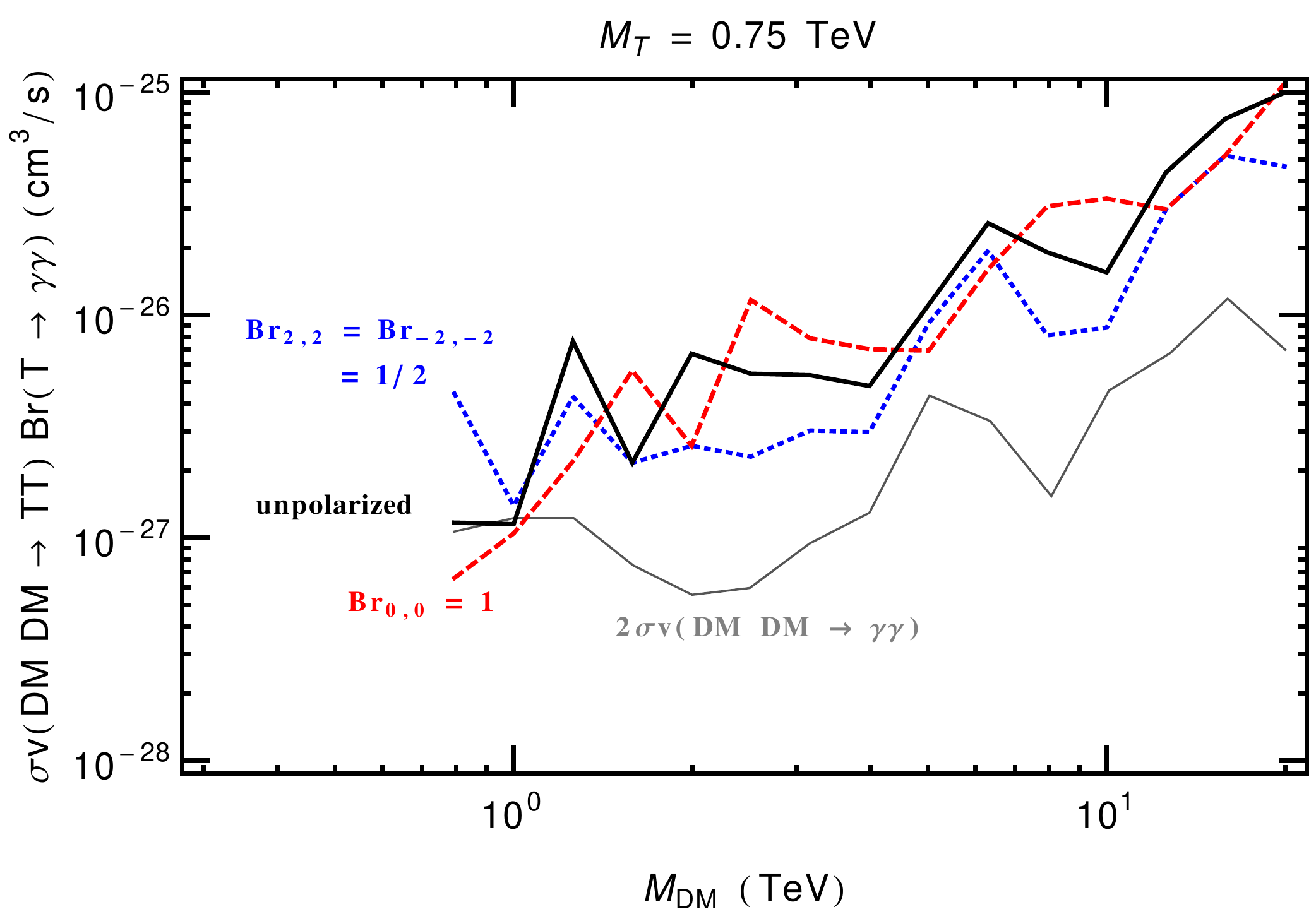}
\includegraphics[width=0.45\textwidth]{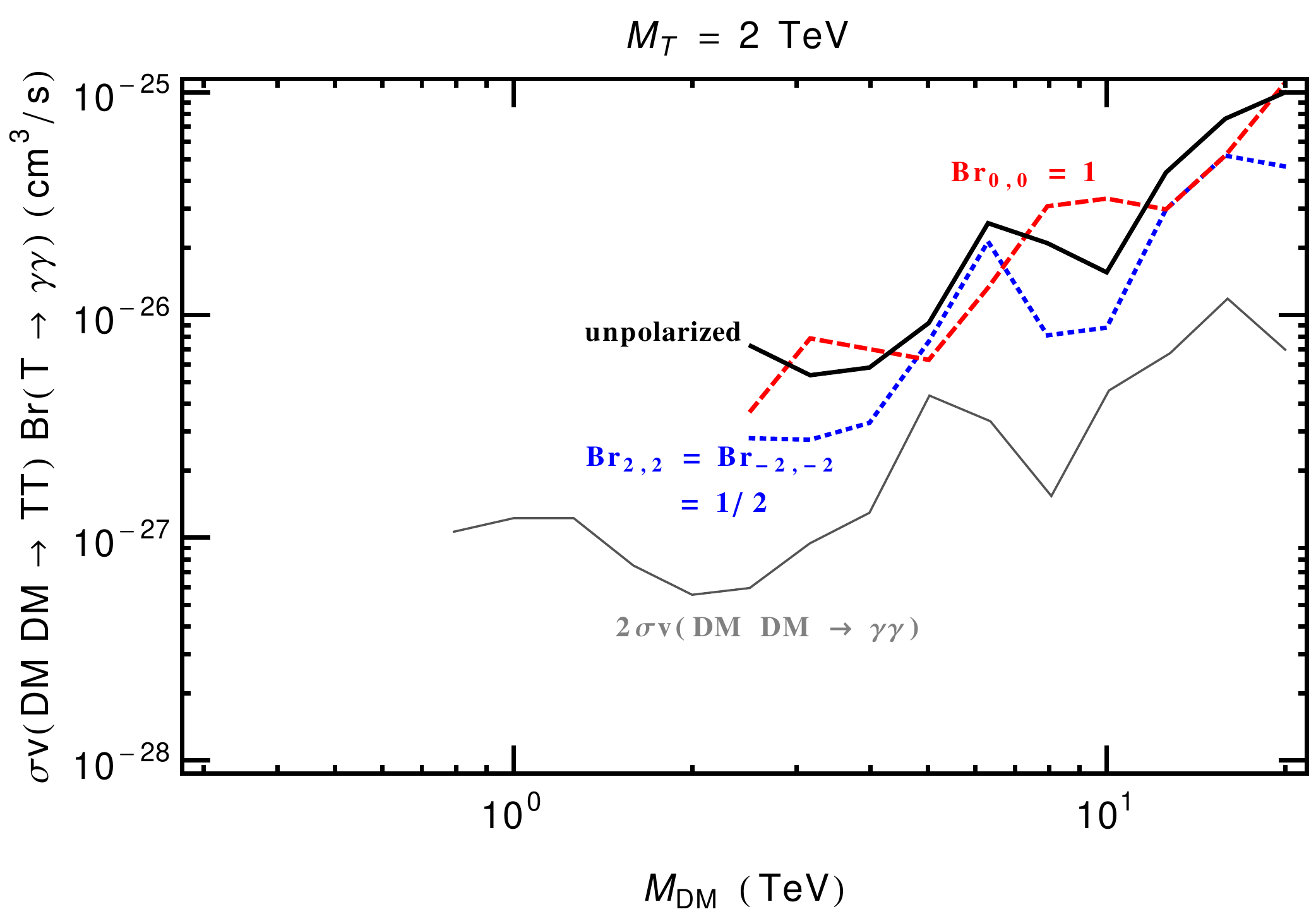}
\caption{ 
95\%~C.L.~upper limits on the cross section $\sigma v(\text{DM DM} \to T T) \Br(T\to\gamma\gamma)$ from the non-observation of gamma-ray spectral features by H.E.S.S., for $M_T = \unit[0.75]{TeV}$ (left) and $M_T=\unit[2]{TeV}$ (right). The limits were derived assuming the Einasto DM distribution profile. For comparison, we also show the corresponding limits for monochromatic gamma-ray lines~\cite{Abramowski:2013ax}.}
\label{fig:HESSLimits}
\end{figure*}

\begin{figure}[t]
\includegraphics[width=0.450\textwidth]{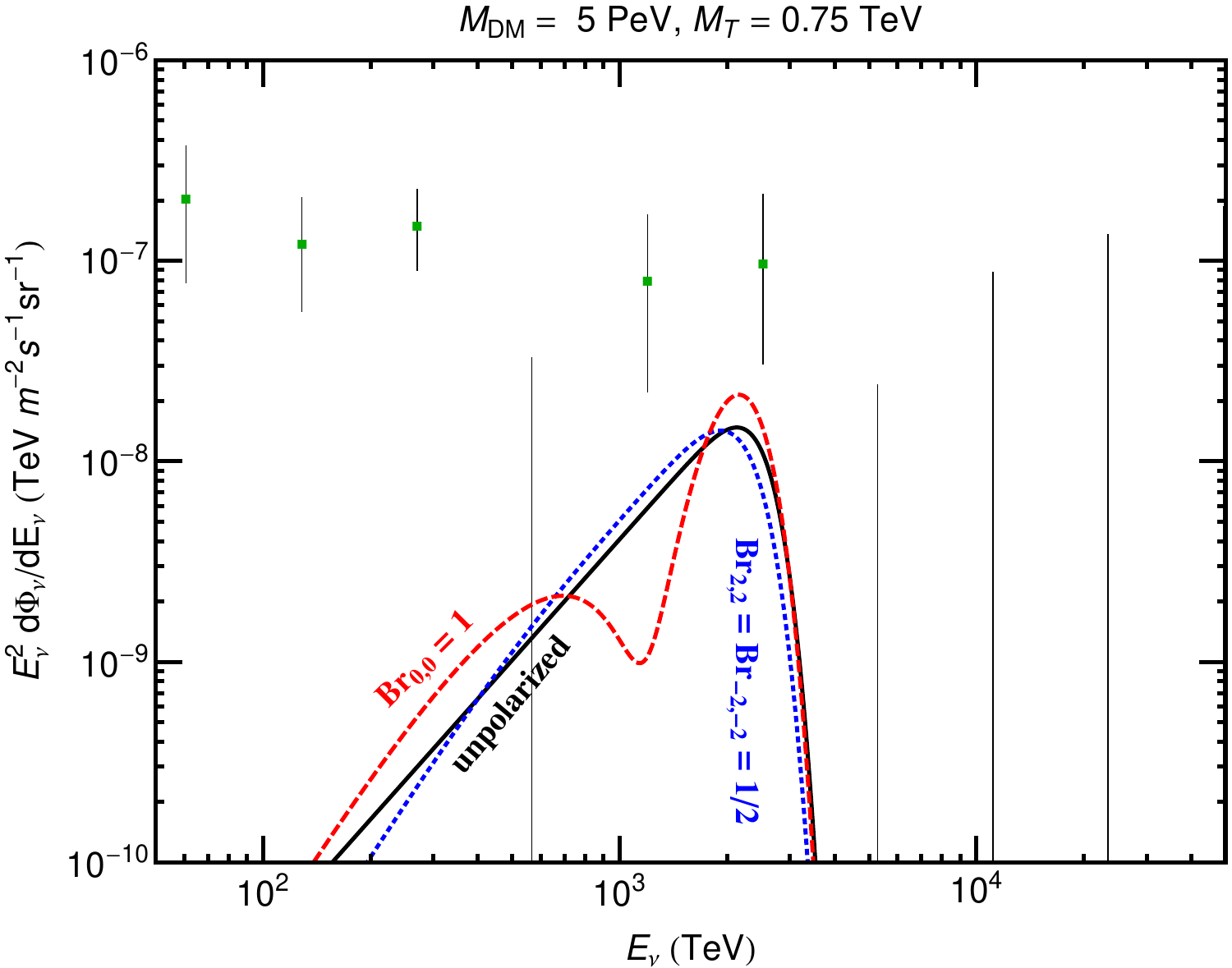}
\caption{ Neutrino signal expected from DM decay into a pair of spin-2 particles $T$ subsequently decaying into $\nu\bar\nu$, along with the corresponding IceCube data~\cite{Aartsen:2014gkd}. For purely illustrative purposes we choose $\Phi_\nu =   \unit[10^{-11}]{m^{-2}s^{-1}sr^{-1}} $ for the overall DM signal and an energy resolution of $15\%$. The unpolarized spectrum is the same as the flat box of a spin-0 particle. 
}
\label{fig:IceCube}
\end{figure}

To compare with the astrophysical background, and also in order to provide a somewhat realistic example, we consider the case of a spin-2 particle with mass $M_T = \unit[750]{GeV}$, DM with $M_\text{DM}=\unit[5]{TeV}$, and assume an annihilation gamma-ray flux of $\Phi_\gamma = \unit[3.5\times 10^{-6}]{m^{-2}s^{-1}sr^{-1}} $ as input in Eq.~\eqref{eq:Aflux}. 
The relation between the total flux and the cross section is given by
\begin{align}
\Phi_\gamma= \frac{ \sigma v (\text{DM DM} \to  T T)\times 4 \,\Br (T\to \gamma\gamma)}{8\pi M_\text{DM}^2} \, \bar{J}_\text{ann} \,,
\label{eq:gammaflux}
\end{align}
where the factor of $4\,\Br (\gamma\gamma) = 2 \,\Br ( \gamma\gamma) [1-\Br ( \gamma\gamma)] + 2 \,[1-\Br (\gamma\gamma)]\Br (\gamma\gamma) + 4\, \Br (\gamma\gamma)^2$ takes into account all possible $T$ decays that lead to photons.
Together with the astrophysical factor $\bar{J}= \unit[7.44 \times 10^{24}]{GeV^2/cm^{5}/sr}$ 
(associated to the Einasto DM distribution profile  for the galactic halo region as specified in Refs.~\cite{Abramowski:2011hc,Abramowski:2013ax}), the flux $\Phi_\gamma = \unit[3.5\times 10^{-6}]{m^{-2}s^{-1}sr^{-1}}$ then corresponds to a cross section $\sigma v (\text{DM DM} \to  T T)\times\Br (T\to \gamma\gamma) = \unit[0.75\times 10^{-26}]{cm^3/s}$.
We choose these values only for illustration purpose; they have no physical relevance as here we do not aim at estimating the relic density or the branching ratio into diphoton final states.
Notice nevertheless that such values are achievable in various TeV-DM models, since many of them give rise to cross sections  larger than canonical thermal value due to non-perturbative effects (for concrete examples see e.g.~Ref.~\cite{Garcia-Cely:2015dda}). 
The resulting differential flux is shown in Fig.~\ref{fig:HESS} (left), compared against the gamma-ray flux from the galactic halo region as measured by the H.E.S.S.~telescope~\cite{Abramowski:2013ax}. As done by the collaboration itself, we multiply the fluxes by $E_\gamma^{2.7}$ so that the spectral features are more visible compared to the essentially flat background. 

All three spectra are distinguishable from the astrophysical background due to their sharp feature. This allows us to calculate the 95\% C.L.~upper limits on the annihilation cross section $\sigma v(\text{DM DM} \to T T) \Br(T\to\gamma\gamma)$ for arbitrary DM masses in the range of H.E.S.S., but still taking $M_T =\unit[750]{GeV}$. In addition we also present the limits for $M_T = \unit[2]{TeV}$. We follow the method pursued by the H.E.S.S.~collaboration in Ref.~\cite{Abramowski:2013ax}, which adopts a phenomenological background model defined by seven parameters.  Our results are reported in Fig.~\ref{fig:HESSLimits} for the cases in which the tensor is  produced unpolarized as well as for $\Br_{2,2}= \Br_{-2,-2}=1/2$ and $\Br_{0,0}=1$.  For the sake of comparison, we also show the corresponding limits for $2\,\sigma v (\text{DM DM}\to \gamma \gamma)$~\cite{Abramowski:2013ax}, which corresponds to monochromatic lines with the same total flux as the other cases. Up to a factor of a few, the limits from polynomial gamma-ray boxes are competitive with the ones from lines. 

As is clear from Figs.~\ref{fig:HESS} and~\ref{fig:HESSLimits}, it is difficult to distinguish the spectra associated to different the polarizations of the spin-2 particle. Doing so crucially relies on the detector resolution. In order to illustrate  this, we show the optimistic case of a 5\% energy resolution in Fig.~\ref{fig:HESS} (right), potentially achievable in future gamma-ray telescopes such as C.T.A.~\cite{Bernlohr:2012we}. We observe that the case $\Br_{2,2} = \Br_{-2,-2} = 1/2$ now gives a narrower feature compared to the other two as well as a lower-energy ``shoulder''. We leave a detailed quantitative comparison for future work. Nevertheless, we would like to remark that the sensitivity of C.T.A.~to flat gamma-ray boxes has already been discussed in Ref.~\cite{Ibarra:2015tya}.

It is straightforward to repeat the above analysis for neutrino spectra. Compared to gamma-ray telescopes, IceCube is sensitive to much higher energies, with several PeV neutrinos detected in the last years~\cite{Aartsen:2014gkd}. Neutrinos with such high energies are more easily obtained from DM \emph{decays}, which require a slight modification of our definition of the total flux from Eqs.~\eqref{eq:neutrinoflux}, but are still described by the spectral function of Eq.~\eqref{eq:Aflux}. Denoting the DM lifetime by $\tau_\text{DM}$, we have
\begin{align}
\Phi_\nu = \frac{ \bar J_\text{dec} }{4\pi M_\text{DM}}\, \frac{4\,\Br (\text{DM}\to TT)\,\Br (T\to \nu\bar\nu)}{\tau_\text{DM}} \,,
\end{align}
where $\bar J_\text{dec}$ is Eq.~\eqref{eq:Jfactor} with $\rho_\text{DM}^2\to\rho_\text{DM}$ and we already assumed neutrinos and anti-neutrinos to be indistinguishable, hence the factor of 4 in the numerator. 
Taking the same benchmark values for $\Br_m$ as above for the gamma-ray flux, we obtain the spectra of Fig.~\ref{fig:IceCube}, using,  for illustration purposes, $\Phi_\nu =   \unit[10^{-11}]{m^{-2}s^{-1}sr^{-1}} $ and $M_\text{DM}=\unit[5]{PeV}$. These spectra are insensitive to the tensor mass as long as $M_T \ll M_\text{DM}$.

We observe that the double-peak shape in the $\Br_{0,0}=1$ spectrum is more pronounced compared to the double-peak photon spectrum (Fig.~\ref{fig:HESS}), which is in part due to the steeper background for gamma rays compared to neutrinos -- which translates into our choices of the vertical axes -- and in part because the actual dip is deeper (see Figs.~\ref{fig:DMDMtoTT}). The main effect is however the same: one of the spectra appears more narrow than the other two, and will lead to DM bounds that are in between those of a box and a monochromatic line (see recent analysis in Ref.~\cite{Aisati:2015vma}).

\section{Connection to the 750~GeV resonance}
\label{sec:diphoton} 

Let us finally turn to the tantalizing LHC diphoton excess~\cite{CMS:2015dxe,CMS:2016owr,atlas_diphoton,ATLAS_diphoton_moriond}. Its simplest explanation assumes a spin-0 or spin-2 particle $R$ of mass $\unit[750]{GeV}$ coupled to photons and protons (spin 1 being excluded by the Landau--Yang theorem~\cite{Landau,Yang:1950rg}). There is not enough data yet to establish the total width of the resonance or its various branching ratios. If this new particle $R$ also couples to DM and can kinematically be produced on shell via annihilations (or decay), we can apply our formalism from the previous sections to determine the resulting indirect detection signatures. The cleanest channel, of course, comes from the monochromatic photons produced in the process $\text{DM DM}\to R^* \to\gamma\gamma$. 
For scalar or vector DM, this annihilation can take place via the $s$ wave. In contrast, for Majorana DM, it is $p$-wave suppressed if $R$ is a real scalar or spin-2 tensor, but $s$ wave for a pseudoscalar $R$. 

If $M_\text{DM} >\unit[750]{GeV}$ and the coupling of $R$ to DM is larger than to photons, the process $\text{DM DM}\to R R$ can be the dominating annihilation channel. The subsequent decay $R\to \gamma\gamma$ (or $R\to\nu\bar\nu$ depending on the model) will then lead to box-shaped spectra as discussed in the previous sections (Fig.~\ref{fig:DMDMtoTT}). In particular, a spin-0 $R$ (or an unpolarized spin-2 $R$) would lead to a \emph{flat}-box spectrum~\cite{Chu:2012qy,Choi:2016cic}. Since a spin-2 particle is generically produced \emph{polarized}, the gamma-ray spectra from a spin-2 diphoton resonance will not be flat and can potentially be distinguished from the spin-0 spectra. The relevant features compared to the power-law background are shown in Fig.~\ref{fig:HESS} (``unpolarized'' being equivalent to a spin-0 resonance).
With the assumed 15\% energy resolution it will be difficult to distinguish the spin-0 and spin-2 case or the different spin-2 polarizations. A confirmation of the spin-2 nature of the diphoton resonance at the LHC, however, would warrant a more detailed quantitative analysis of the \emph{differentiation} between the different polarizations in indirect detection. 
Nevertheless, the presence of the spectral features already allows us to set limits on the DM annihilation cross section into the diphoton resonance, as shown in Fig.~\ref{fig:HESSLimits}.

Seeing as a universal coupling of the $\unit[750]{GeV}$ spin-2 particle $R$ is disfavored by non-observation of a dilepton resonance~\cite{Franceschini:2015kwy,Low:2015qep,Giddings:2016sfr}, it is reasonable to assume that the coupling (and branching ratio) of $R$ to neutrinos is smaller than to photons. The neutrino spectra can nevertheless be of interest because neutrino telescopes such as IceCube can probe neutrino energies up to PeV, whereas H.E.S.S.~only goes up to $\unit[20]{TeV}$.\footnote{Since DM annihilations and decays unavoidably also lead to a continuous gamma-ray spectrum at lower energies~\cite{Ciafaloni:2010ti}, additional constraints can be derived~\cite{Profumo:2016idl,Queiroz:2016zwd}.}
The neutrino spectra derived here could for example be used to explain the ``peak'' of three PeV neutrinos in IceCube~\cite{Aartsen:2014gkd,Vincent:2016nut} by setting $M_\text{DM}\sim \unit{PeV}$ and e.g.~$M_T =\unit[750]{GeV}$, analogous to other DM-inspired explanations~\cite{Feldstein:2013kka,Esmaili:2013gha,Rott:2014kfa} (see Fig.~\ref{fig:IceCube}). Astrophysical explanations for this statistically insignificant observation are readily available; in particular, the highest PeV neutrino event appears to be in temporal and positional coincidence with a blazar outburst~\cite{Kadler:2016ygj}. We thus omit a more detailed discussion. 

An interesting point to end this section on is the model-independent correlation of neutrino and gamma-ray spectra (Fig.~\ref{fig:DMDMtoTT}), which should in principle help to determine the polarization of the mediator once signals are observed both by neutrino and Cherenkov telescopes.

\section{Conclusion}
\label{sec:conclusion}

We have investigated DM annihilations (and decays) into two particles with arbitrary spin, which subsequently undergo two-body decay into photons or neutrinos. The resulting differential flux of energy is a box spectrum with an overall polynomial shape given by Eq.~\eqref{eq:Aflux}. This formula can be applied to an arbitrary model. 

The relative weight of each polynomial is determined by the model-dependent coefficients $\Br_m$ and $C_{m}$. The former denotes the relative production probability of $X$ with polarization $m$, whereas the latter describes the angular distribution of the decay products of $X$, e.g.~photons or neutrinos, when their helicity difference is $m$. Only if \emph{both} $\Br_m$ and $C_m$ actually depend on $m$ do we obtain a polynomial shape for the spectrum instead of a flat box.
We have argued that this is generically the case if the intermediary particle $X$ has spin one or two.

In the case of DM annihilating into massive gauge bosons, the dependence of  $\Br_m$ or $C_m$ on the angular momentum $m$ is rooted in the GBET. Similarly, for DM annihilating into spin-2 particles coupled to the DM energy--momentum tensor, the states with helicity $m=\pm1$ decouple for heavy DM masses. Moreover, branching ratios into states with helicity $m=0$ dominate unless some selection rule forbids them. This happens for Majorana DM annihilations, in which case the branching ratios into $m=\pm2$ dominate, yielding highly non-trivial spectra for photons and neutrinos.

We have discussed the implications of these effects for non-relativistic DM annihilations or decays into the hypothetical \unit[750]{GeV} resonance responsible for the diphoton excess observed at the LHC. 
As an example, we have derived the gamma-ray spectrum from DM annihilations in Fig.~\ref{fig:HESS} for a particular benchmark mass at the TeV scale and compared it against  the astrophysical background as measured by the H.E.S.S.~telescope. Similarly, we have calculated the neutrino spectrum of decaying PeV DM in Fig.~\ref{fig:IceCube} and compared it against the neutrino flux measured by IceCube.
In this discussion, we assume that spin-2 particles couple to the energy--momentum tensor. 

Finally, by exploiting the gamma-ray spectral features that are produced when DM annihilates into the scalar/tensor particle, we have also derived an upper limit on the corresponding annihilation cross section. The results are shown in Fig.~\ref{fig:HESSLimits}. With the energy resolution of current gamma-ray telescopes it is not yet feasible to distinguish all the possible spectral shapes. Nevertheless, that is not necessarily an obstacle for future gamma-ray or neutrino telescopes, which could in principle resolve the different  polynomials and determine the polarization of the mediator particle. A quantitative analysis is left for future work.

\section*{Acknowledgements}
We thank Angnis Schmidt-May for insights into massive spin-2 particles and Cha{\"i}mae El Aisati for discussions on neutrino spectra and for carefully reading the manuscript.
CGC is supported by the IISN and the Belgian Federal Science Policy through the Interuniversity Attraction Pole P7/37 ``Fundamental Interactions''.
JH is a postdoctoral researcher of the F.R.S.-FNRS.
We acknowledge the use of \texttt{Package-X}~\cite{Patel:2015tea} and \texttt{JaxoDraw}~\cite{Binosi:2003yf}.

\appendix

\begin{widetext}
\section{Cross section formulae}

For the process $VV\to ST$ (Eq.~\eqref{eq:VVtoST}), the kinematic function $u_{VVST}$ and the helicity branching ratios are given by
\begin{align}
\begin{split}
u_{VVST} (x,y) &\equiv \frac{\sqrt{x^2+(4-y)^2-2 x (4+y)}}{176 (4-y)^2 (4-x-y)^2} \left[x^4 \left(44-12 y+y^2\right)+(4-y)^4 \left(44-12 y+y^2\right)\right.\\
&\quad -4 x^3 \left(-144+156 y-28 y^2+y^3\right)-4 x (4-y)^2 \left(-144+156 y-28 y^2+y^3\right)\\
&\quad \left.+2 x^2 \left(12352-6624 y+1364 y^2-108 y^3+3 y^4\right)\right] ,
\end{split}
\label{eq:uVVST}\\
\Br_2/\Br_1 &= \frac{16 r_T^2}{(4-r_S^2+r_T^2)^2} \,,
\label{eq:BrVVST1}\\
\begin{split}
\Br_2/\Br_0 &=384 \left(4-r_S^2\right)^2 r_T^4\left[\left(4-r_S^2\right)^4 \left(44-12 r_S^2+r_S^4\right)-4 \left(4-r_S^2\right)^2 \left(48+60 r_S^2-16 r_S^4+r_S^6\right) r_T^2\right.\\
&\quad +2 \left(3136-1248 r_S^2+404 r_S^4-60 r_S^6+3 r_S^8\right) r_T^4-4 \left(48+60 r_S^2-16 r_S^4+r_S^6\right) r_T^6\\
&\quad \left.+\left(44-12 r_S^2+r_S^4\right) r_T^8\right]^{-1} .
\end{split}
\label{eq:BrVVST2}
\end{align}
For $F\bar F\to VT$ (Eq.~\eqref{eq:FFtoVT}), the kinematic function $u_{FFVT}$ and the helicity branching ratios are given by
\begin{align}
\begin{split}
u_{FFVT} (x,y) &\equiv \frac{\sqrt{x^2+(4-y)^2-2 x (4+y)}}{32 (4-y)^2 (4-x-y)^2} \left[(4-y)^4 (8+y)+x^4 (48+y)+2 x (4-y)^2 \left(16-64 y+3 y^2\right)\right.\\
&\quad \left.+x^3 \left(-288+272 y+6 y^2\right)+2 x^2 \left(64+112 y+92 y^2-7 y^3\right)\right] ,
\label{eq:uFFVT}
\end{split}\\
\begin{split}
\Br_2/\Br_1 &=\frac{8 r_T^2 \left(4-r_T^2-r_V^2\right)^2}{\left(4-r_V^2\right)^4-2 r_T^2 \left(4-r_V^2\right)^2 \left(4+r_V^2\right)+r_T^4 \left(16+24 r_V^2+r_V^4\right)} \,,
\label{eq:BrFFVT1}
\end{split}\\
\begin{split}
\Br_2/\Br_0 &=\frac{24 r_T^4 \left(4-r_T^2-r_V^2\right)^2}{32 r_T^6 r_V^2+r_T^8 r_V^2+\left(4-r_V^2\right)^4 \left(8+r_V^2\right)-16 r_T^2 \left(4-r_V^2\right)^2 \left(4+5 r_V^2\right)+2 r_T^4 \left(4+r_V^2\right) \left(16+48 r_V^2-r_V^4\right)} \,.
\label{eq:BrFFVT2}
\end{split}
\end{align}
\\
\end{widetext}

\bibliographystyle{utcaps_mod}
\bibliography{BIB}
\end{document}